\documentclass[anonymous]{llncs}

\usepackage{setspace}
\usepackage{bm}
\usepackage{xcolor}
\usepackage{booktabs}
\usepackage{microtype}
\usepackage{amsfonts}
\usepackage{xspace}
\usepackage{wasysym}
\usepackage{enumitem,etoolbox}
\usepackage{array}
\usepackage{microtype}
\usepackage{multirow}
\usepackage{framed}
\usepackage{caption}
\usepackage{threeparttable}
\usepackage{appendix}
\usepackage[labelformat=simple]{subcaption}

\usepackage[ruled,vlined]{algorithm2e}

\SetCommentSty{mycommfont}

\SetKwInOut{Input}{Input}
\SetKwInOut{Output}{Output}
\SetKwInOut{Except}{Except.}
\SetKw{None}{nil}
\SetKw{Cont}{continue}
\SetKw{Break}{break}
\SetKwProg{Fn}{Function}{:}{}
\SetKwFunction{Initp}{InitPattern}
\SetKwFunction{Nextp}{NextPattern}
\SetKwFunction{Initss}{InitSerSched}
\SetKwFunction{Nextss}{NextSerSched}
\SetAlCapFnt{\small}
\SetAlCapNameFnt{\small}
\LinesNumbered
\DontPrintSemicolon

\usepackage{tikz}
\usetikzlibrary{positioning}
\tikzstyle{period} = [draw=white, fill=gray!30, thick,
    rectangle, rounded corners, minimum size=9ex]
\usepackage{subcaption}
\usepackage{soul}
\AtBeginEnvironment{verbatim}{\setlist[trivlist]{nosep}}
\usepackage{flushend}
\usepackage{balance}
\usepackage{graphicx}

\usepackage{amsthm}
\usepackage{amsmath}
\usepackage{amssymb}

\usepackage{bbding}
\usepackage{array}
\usepackage{multirow}
\usepackage{listings}
\usepackage{color}
\usepackage{tcolorbox}
\usepackage{xcolor,colortbl}
\usepackage{enumitem}
\usepackage{adjustbox}
\usepackage[normalem]{ulem}
\usepackage{pifont}
\usepackage[colorlinks=true,linkcolor=blue,anchorcolor=red,citecolor=green]{hyperref}

\usetikzlibrary{shapes,positioning}
\makeatletter
\let\old@lstKV@SwitchCases\lstKV@SwitchCases
\def\lstKV@SwitchCases#1#2#3{}
\makeatother
\usepackage{lstlinebgrd}
\makeatletter
\let\lstKV@SwitchCases\old@lstKV@SwitchCases

\lst@Key{numbers}{none}{%
    \def\lst@PlaceNumber{\lst@linebgrd}%
    \lstKV@SwitchCases{#1}%
    {none:\\%
        left:\def\lst@PlaceNumber{\llap{\normalfont
                \lst@numberstyle{\thelstnumber}\kern\lst@numbersep}\lst@linebgrd}\\%
        right:\def\lst@PlaceNumber{\rlap{\normalfont
                \kern\linewidth \kern\lst@numbersep
                \lst@numberstyle{\thelstnumber}}\lst@linebgrd}%
    }{\PackageError{Listings}{Numbers #1 unknown}\@ehc}}
\makeatother

\definecolor{rowcolor}{HTML}{ECEFF4}

\setlist[itemize]{leftmargin=*}

\definecolor{dkgreen}{rgb}{0,0.6,0}
\definecolor{gray}{rgb}{0.5,0.5,0.5}
\definecolor{mauve}{rgb}{0.58,0,0.82}

\newcommand{\eg}{\hbox{\emph{e.g.}}\xspace}
\newcommand{\ie}{\hbox{\emph{i.e.}}\xspace}
\newcommand{\etc}{\hbox{\emph{etc.}}\xspace}

\newcommand{\name}{\textsc{AutoSpec}\xspace}
\newcommand{\framac}{\textsc{Frama-C}\xspace}
\newcommand{\wpplugin}{\textsc{WP}\xspace}
\setlist[itemize]{leftmargin=*}
\setlist[enumerate]{leftmargin=*}
\newlist{steps}{enumerate}{1}
\setlist[steps, 1]{label = \textbf{RQ\arabic*.}}

\newcommand{\wc}[1]{\textcolor{blue}{{[wc: #1]}}}
\newcommand{\xzw}[1]{\textcolor{red}{{[xzw: #1]}}}

\newcommand{\qin}[1]{\textcolor{orange}{{[qin: #1]}}}

\begin{document}

%
% The "title" command has an optional parameter, allowing the author to define a "short title" to be used in page headers.
\title{Enchanting Program Specification Synthesis by Large Language Models using Static Analysis and Program Verification}
%\title{Step-by-Step and Iterative Program Specification Generation with Large Language Models for Automating Deductive Verification}
\author{Cheng Wen\inst{1} \and Jialun Cao\inst{2,}\thanks{Corresponding authors: Jialun Cao and Shengchao Qin.} \and Jie Su\inst{1} \and Zhiwu Xu\inst{3} \and Shengchao Qin\inst{4,{\star}} \and Mengda He\inst{4} \and Haokun Li\inst{4} \and Shing-Chi Cheung\inst{2} \and Cong Tian\inst{5}}
\institute{Guangzhou Institute of Technology, Xidian University, Guangzhou, China \and The Hong Kong University of Science and Technology, Hong Kong, China \and College of Computer Science and Software Engineering, Shenzhen University, China \and Fermat Labs, Huawei, Hong Kong, China \and ICTT and ISN Laboratory, Xidian University, Xi'an, China \\
}

\maketitle

%
% By default, the full list of authors will be used in the page headers. Often, this list is too long, and will overlap
% other information printed in the page headers. This command allows the author to define a more concise list
% of authors' names for this purpose.
%\renewcommand{\shortauthors}{Cheng Wen et al.}

% Abstract
\begin{abstract}
Formal verification provides a rigorous and systematic approach to ensure the correctness and reliability of software systems. 
Yet, constructing specifications for the full proof relies on domain expertise and non-trivial manpower. In view of such needs, an automated approach for specification synthesis is desired.
\iffalse\qin{The following 2 sentences require repurchasing as there are works that can handle spec inference for highly-involved heap-allocated data structures eg. CAV14:Shape Analysis via Second-Order Bi-Abduction. There are many works around invariant generation in CAV or PL conferences. The following is a first attempt to rephrase the 1st sentence.}\fi
While existing automated \iffalse specification synthesis\fi approaches are limited in their versatility, \ie,  they either focus only on synthesizing loop invariants for numerical programs\iffalse with one or two loops over integer variables\fi\iffalse\qin{ADD CITATIONS in Intro}\fi, or are tailored for specific types of programs or invariants.
\iffalse\qin{ADD CITATIONs in Intro: Naijun's work, Quang Loc Le: CAV14, Ton Chan Le: PLDI15, Sankaranarayanan: POPL04, etc}. \qin{We need to rephrase the next sentence to point out that what we can deal with in this work is more general than existing works.}\fi
Programs involving multiple complicated data types (\eg, arrays, pointers) and code structures (\eg, nested loops, function calls) are often beyond their capabilities. 
\iffalse Also, loop invariants alone can hardly lead to a full proof, other specifications such as preconditions and postconditions are also indispensable for the verification.\fi 
\iffalse Moreover, for full proof, other specifications such as preconditions and postconditions are also indispensable for verification.\fi
To help bridge this gap, we present \name, an automated approach to synthesize specifications \iffalse synthesis \fi for automated program verification. 
It overcomes the shortcomings of existing work in specification versatility, synthesizing satisfiable and adequate specifications for full proof.
It is driven by static analysis and program verification, and is empowered by large language models (LLMs).
{\name} addresses the practical challenges \iffalse three-fold\fi in three ways: (1) driving \name by static analysis and program verification, LLMs serve as generators to generate candidate specifications, 
(2) \iffalse decomposing the programs to lead LLMs' attention\fi programs are decomposed to direct the attention of LLMs, 
and (3) \iffalse validating the candidate specifications\fi candidate specifications are validated in each round to avoid error accumulation during the interaction with LLMs. 
\iffalse By doing so\fi In this way, {\name} can incrementally and iteratively generate satisfiable and adequate specifications. 
The evaluation shows its effectiveness and usefulness, as it outperforms existing works by successfully verifying 79\% of programs through automatic specification synthesis, a significant improvement of 1.592x.
It can also be successfully applied to verify the programs in a real-world X509-parser project. 
\end{abstract}


%
% Keywords. The author(s) should pick words that accurately describe the work being
% presented. Separate the keywords with commas.

%\keywords{Program Specification, Verification, Hoare logic, Large Language Model}

%
% This command processes the author and affiliation and title information and builds
% the first part of the formatted document.
%\IEEEpeerreviewmaketitle
%\settopmatter{printfolios=true}
%\maketitle

%\hspace*{\fill} \\
%\hspace*{\fill} \\
%\hspace*{\fill} \\
%{\color{gray}This is the author's version of the work. It is posted here for your personal use. Not for redistribution. The definitive version was published in the proceedings of XXX.}

%\newpage
\section{Introduction}\label{sec:intro}
Program verification offers a rigorous way to assuring the important properties of a program. Its automation, however, needs to address the challenge of proof construction~\cite{hahnle2019deductive,code2inv18}. 
Domain expertise is {\iffalse often\wc{space issue}\fi}required for non-trivial proof construction, where human experts identify important program properties, write the \textit{specifications}
(\eg, the pre/post-conditions, invariants, and contracts written in certain \textit{specification languages}), and then use these specifications to prove the properties.

Despite the immense demand for software verification in the industry~\cite{ebalard2019journey,efremov2018deductive,dordowsky2015experimental,blanchard2015case,kosmatov2014case}, \textit{\textbf{manual verification by experts remains the primary approach in practice.}}
To reduce human effort, \textbf{\textit{automated specification synthesis}} is desired. 
Ideally, given a program and a property \iffalse under verification\fi to be verified, we expect {the specifications that are sufficient for a full proof could be synthesized automatically}.

{\textbf{Research gap -- }}Prior works are limited in \iffalse their\wc{space issue}\fi versatility, \ie, the ability to simultaneously handle \textit{different types of specifications} (\eg, invariants, preconditions, postconditions), \textit{code structures} (\eg, multiple function calls, multiple/nested loops), and \textit{data structures} (\eg, arrays, pointers), leaving room for improvement towards achieving full automation in proof construction. 
\iffalse In particular, a branch of\fi Existing works focus only on loop invariants~\cite{dillig2013inductive,lin2021inferring,yu2023loop}, preconditions~\cite{cousot2013automatic,padhi2016data}, or postconditions~\cite{popeea2006inferring,su2018using,singleton2018algorithm}. Moreover, most works on loop invariant synthesis can only handle{\iffalse programs with one loop with integer variables\fi} numerical programs~\cite{ryan2020cln2inv,code2inv18,yao2020learning,gupta2009invgen} or are tailored for specific types of programs or invariants~\cite{le2014shape,wang2021synthesizing,feng2017finding,gan2020nonlinear,le2015termination,vazquez2020maximum}\iffalse\qin{add citations}\fi. To handle various types of specifications simultaneously and to process programs with various code and data structures, a versatile approach is required.

\textbf{{Challenges --}} Although the use of large language models (LLMs) such as ChatGPT may provide a straightforward solution to program specification generation, it is not a panacea. The generated specifications are mostly incorrect due to three intrinsic weaknesses of LLMs.
First, \textit{\textbf{LLMs can make mistakes.}}
Even for the well-trained programming language Python, ChatGPT-4 and ChatGPT-3.5 only achieve 67.0\% and 48.1\% accuracy in program synthesis~\cite{gpt4report}. In comparison with programming languages\iffalse (\eg Python)\fi, LLMs are much less trained in specification languages. Therefore, LLMs generally perform worse in synthesizing specifications than programs.
Since the generated specifications are error-prone, we need an effective technique to detect incorrect specifications, which are meaningless to verify. 
Second, \textit{\textbf{LLMs may not attend to the tokens we want them to}}. Self-attention may pay no, less, or wrong attention to the tokens that we want it to. Recent research even pointed out a phenomenon called ``lost in the middle''~\cite{liu2023lost}, observing that LLMs pay little attention to the middle if the context goes extra long. In our case, the synthesized specifications are desired to capture and describe as many program behaviors as possible. Directly adopting the holistic synthesis (\ie, synthesizing all specifications at once) may yield unsatisfactory outcomes. 
Third, \textit{\textbf{errors accumulate in the output of LLMs.}} LLMs are auto-regressive. If they make mistakes, these wrong outputs get added to their inputs in the next round, leading to way more wrong outputs. It lays a hidden risk when taking advantage of LLMs' dialogue features, especially in an incremental manner (\ie, incrementally synthesizing specifications based on previously generated ones).

\textbf{Insight --} To address the above challenges, \textbf{\textit{our key insight is to let static analysis and program verification take the lead, while hiring LLMs to synthesize candidate specifications.}} 
Static analysis parses a given program into pieces, and \iffalse sporting\fi passes each program piece \iffalse of code\fi in turn to LLMs by inserting a placeholder in \iffalse the program\fi  it. 
Paying attention to the spotted part, LLMs generate a list of specifications as candidates.
Subsequently, a theorem prover validates the generated specifications and keeps the validated ones in the next round of synthesis. The iteration process terminates when the property under verification has been proved, or the iteration reaches a predefined limit.

\textbf{{Solution --}} Bearing the insight, we present {\name}, an LLM-empowered framework for generating specifications. It tackles the three above-mentioned limitations of directly adopting LLM in three perspectives.
First, \textbf{\textit{it decomposes the program hierarchically}} and employs LLMs to generate specifications incrementally in a bottom-up manner. 
This allows LLMs to focus on a selected part of the program and generate specifications only for the selected context. Thus, the limitation of context fragmentation could be largely alleviated. 
Second, \textit{\textbf{it validates the generated specifications}} using theorem provers.
Specifications that are inconsistent with programs' behaviors and contradict the properties under verification will be discarded. This post-process ensures that the generated specifications are satisfiable by the source code and the properties under verification.
\iffalse the consistency between the source code and specifications, as well as the satisfiability of properties under verification.\fi
Third, \textit{\textbf{it iteratively enhances the specifications}} by employing LLMs to generate more specifications until they are adequate to verify the properties under verification \iffalse sufficient specifications are generated\fi or the number of iterations reaches the predefined upper bound.

We evaluate the effectiveness of \name by conducting experiments on 251 C programs across four benchmarks, each with specific properties to be verified.
We compare \name with three state-of-the-art approaches: Pilat, Code2inv, and CLN2Inv.
The result shows \name can successfully handle 79\%\,(=\,199\,/\,251) programs with various structures (\eg, linear/multiple/nested loops, arrays, pointers), while existing approaches can only handle programs with linear loops. As a result, 59.2\%\,(=\,(199\,-\,125)\,/\,125) more programs can be successfully handled by \name. 
% with 29\% (= 1 - 125 / 251) programs unsolved. 
The result also shows that \name outperforms these approaches regarding effectiveness and expressiveness when accurately inferring program specifications.
To further indicate its usefulness, we apply \name to a real-world X509-parser project, demonstrating its ability to automatically generate satisfiable and adequate specifications for six functions within a few minutes.
In addition, {\iffalse we further analyze the performance of \name using an ablation study that assesses the individual contributions made by different components of \name.\fi}
the ablation study reveals that the program decomposition and the hierarchical specification generation components contribute most to performance improvement.

In summary, this paper makes the following contributions:
\begin{itemize}[leftmargin=*, topsep=2pt, itemsep=1pt]
    \item \textbf{Significance.} We present an automated specification synthesis approach, \name, for program verification. \iffalse It overcomes the shortcomings of existing work in specification versatility, and synthesizes satisfiable and adequate specifications for a full proof.\fi \name is driven by static analysis and program verification, and empowered by LLMs. It can synthesize different types of specifications (\eg,\;invariants, preconditions, postconditions) for programs with various structures (\eg,\;linear/multiple/nested loops, arrays, pointers).
    \item \textbf{Originality.} 
    {\name} tackles the practical challenges{\iffalse associated with\fi} for applying LLMs to specification synthesis: \iffalse driving \name by static analysis and program verification, LLMs serve as candidate specification generators, \fi
    It decomposes \iffalse decomposing\fi the programs hierarchically to lead LLMs' attention, 
    and validates \iffalse validating\fi  the \iffalse candidate \fi  specifications at each round to avoid error accumulation. 
    By doing so, {\name} can incrementally and iteratively generate satisfiable and adequate specifications to verify the desired properties.
    \item \textbf{Usefulness.} We evaluate {\name} on four benchmarks and a real-world X509-parser. The four benchmarks include 251 programs with linear/multiple/nested loops, array structures, pointers, \etc. \name can successfully handle 79\% of them, 1.592x outperforming existing works. The experiment result shows the effectiveness, expressiveness, and generalizability of \name. 
\end{itemize}

\section{Background and Motivation}\label{sec:moti}
% \begin{figure} [t]
%     \centering
%     \vspace{-0ex}
%     \includegraphics[width=0.7\linewidth]{Figure/moti.pdf}
%     \setlength{\abovecaptionskip}{-0pt}
%     \setlength{\belowcaptionskip}{-0pt}
%     \caption{ACSL annotations to functional proof of the bubble sort}
%     \label{fig:moti}
% \end{figure}

% \begin{listing}[ht]
% \inputminted[
% frame=lines,
% framesep=1mm,
% baselinestretch=1.0,
% fontsize=\footnotesize,
% linenos,
% xleftmargin=5.5ex,
% baselinestretch=1.0,
% ]{c}{../Code/moti.c}
% \caption{ACSL annotations to functional proof of the bubble sort}
% \label{fig:moti}
% \end{listing}

% \setlength{\textfloatsep}{10pt}
\definecolor{stubbg}{HTML}{FCF3D5}
\begin{figure}[t]
\begin{lstlisting}[
		language=c,
        basicstyle=\linespread{1.0}\tiny,
        % baselinestretch=1.0,
		% morekeywords={var},
        aboveskip=-30pt,
        belowskip=0pt,
        captionpos=none,
		%caption={ACSL Annotations to Functional Proof of \footnotesize{Bubble Sort}},
		%label={fig:moti},
		escapechar=|,
		% linebackgroundcolor = {\ifnum \value{lstnumber} > 2 \ifnum \value{lstnumber} < 10 \color{stubbg} \fi \fi},
		numbers=left
	]
#include <limits.h>

/*@
requires \valid(a);
requires \valid(b);
ensures *a == \old(*b);
ensures *b == \old(*a);
assigns *a,*b;
*/
void swap(int *a, int *b) {  |\hl{\textbf{ 1. Swap}}| 
  int temp = *a;
  *a = *b;
  *b = temp;
}

/*@ 
requires \valid(array+(0..n-1));
requires 0 < n < INT_MAX;
ensures \forall integer i; 0 < i < n ==> array[i-1] <= array[i];
*/
void bubbleSort(int *array, int n) {  |\hl{\textbf{ 4. bubbleSort}}| 
  if (n <= 0) return;
  int i, j;
  /*@
  loop|\;|invariant|\;|0 <= i < n;
  loop|\;|invariant|\;|\|\!|forall integer k; i <= k < n-1 ==> array[k] <= array[k+1];
  loop|\;|invariant|\;|\|\!|forall integer k; 0 <= k < i+1 <= n-1 ==> array[k] <= array[|\!|i+1|\!|];
  loop assigns i, j, array[0..n-1];
  */
  for(i = n - 1; i > 0; i--) {    |\hl{\textbf{ 3. Outer loop}}| 
    /*@
    loop|\;|invariant|\;|0 <= j <= i < n;
    loop|\;|invariant|\;|\|\!|forall integer k; 0 <= k <= j ==> array[k] <= array[j];
    loop|\;|invariant|\;|\|\!|forall|\;|integer|\;|k; 0 <= k < i+1 <= n-1 ==> array[k] <= array[|\!|i+1|\!|];
    loop assigns j, array[0..i];
    */
    for(j = 0; j < i; j++) {      |\hl{\textbf{ 2. Inner loop}}| 
      if (array[j] > array[j+1]) {
        swap(&array[j], &array[j+1]);
      }
    }
  }
}

void main() {
  int array[5000] = {..., 5, 4, 3, 2, 1};
  bubbleSort(array, 5000);
  //@ assert \forall int i; 0 < i < 5000 ==> array[i-1] <= array[i];
}
\end{lstlisting}
    \setlength{\abovecaptionskip}{5pt}
    \setlength{\belowcaptionskip}{-15pt}
    \caption{ACSL Annotations to Functional Proof of \footnotesize{Bubble Sort}}
    \label{fig:moti}
\end{figure}

Listing~\ref{fig:moti} illustrates a C program that implements the \texttt{bubble sort} (sorting a 5000-element array of integers in ascending order), where the \textit{property to be verified} (line 48) prescribes that after sorting, any index \texttt{i} between 1 and 4999, the element at \texttt{array[i-1]} is no larger than the element at \texttt{array[i]}.
To verify the property, we use a specification language for C programs, ACSL~\cite{baudin2021acsl} (ANSI/ISO-C Specification Language) to write the proof. It appears in the form of code comments (annotated by \texttt{\color{teal}//@ ...} or \texttt{\color{teal}/*@ ... */}) and does not affect the program execution. 
The ACSL-annotated program can be directly fed to \iffalse the \textbf{\textit{theorem prover}}\fi auto-active verification tool (\framac~\cite{frama-c} in this paper) to prove the properties. 

In the running example, specifications in the program prescribe the \textit{preconditions} (begin with \texttt{\color{teal}$\backslash$requires}), \textit{postconditions} (begin with \texttt{\color{teal}$\backslash$ensures}), 
and \textit{loop invariants} (begin with \texttt{\color{teal}loop invariant})\footnote{ACSL has more keywords with rich expressiveness. Refer to the documentation~\cite{baudin2021acsl}.}. 
To prove the property in line 48, practitioners usually write specifications \textit{in a bottom-up manner}, that is, from line 47 tracing to \texttt{bubbleSort} (line\;21), then from line 39, tracing to {\texttt{swap}}\;(line\;10). 
Starting from \textbf{\texttt{swap}}, practitioners identify the inputs and outputs of the \texttt{swap} function and write the pre/post-conditions (lines 4-8). 
In particular, the precondition (lines 4-5) requires the two input pointers to be valid (\ie,\;they can be safely accessed), which is necessary to ensure the safe execution of the operations involving dereferencing. Additionally, the postcondition ensures that the values of \texttt{*a} and \texttt{*b} are swapped (lines 6-7) and assigned (line 8) during execution.

Then tracing back to where \texttt{swap} is called, \ie, inside \texttt{bubbleSort}, it can be challenging because it contains nested loops. In a bottom-up manner, the \textbf{\textit{inner\;loop}} of {{\texttt{bubbleSort}}} (lines 37-41) is first analyzed.
In particular, to verify a loop, it is composed of (1) {loop invariants} (\ie, general conditions that hold before/during/after the loop execution, begin with \texttt{\color{teal}loop invariant}), and possibly (2) the list of assigned variables (begin with \texttt{\color{teal}assign}).
In the example, practitioners analyze the inner loop and write specifications in lines 31-36. Specifically, the index \texttt{j} should fall into the range of 0 to n (line 32), the elements from index \texttt{0} to \texttt{j} are not larger than the element at \texttt{j} (line 33), and all elements from index \texttt{0} to \texttt{i} are smaller than or equal to the element at \texttt{i+1} (line 34). Also, the variables to be assigned in this inner loop include \texttt{j} and first-\texttt{i} elements in the \texttt{array} (line 35). Similarly, for \textbf{\textit{the outer loop}} (lines 30-42), lines 24-29 describe the range of index \texttt{i} (line 25), invariants (lines 26-27) and assigned variables (line 28).

Finally, practitioners analyze \textbf{\texttt{bubbleSort}} (lines 21-43), identifying that the first-\texttt{n} elements of\;\texttt{array}\;can be safely accessed\;(line\;17), \texttt{n} must be greater than zero\;(line\;18). After execution, the array is in ascending order\;(line\;19).
\iffalse \\ \fi
Once all the specifications are written, they are fed into a prover/verification tool), {\framac}~\cite{frama-c} which supports ACSL to verify the \textbf{\textit{satisfiability}} (\ie,\;the specifications satisfy the program) and \textbf{\textit{adequacy}} (\ie,\;the specifications are sufficient to verify the desired properties) of all specifications until the desired property verification succeeds. If the verification fails, practitioners debug and refine the specifications. 

From this example, we can see that the manual efforts to write specifications are non-trivial. Even for a simple algorithm such as bubble sort. 
In practice, the program under verification could be on a far larger scale, which brings a huge workload to practitioners, motivating the \textit{automated specification synthesis}. 

\textbf{Motivation -- }Existing automated specification synthesis works can only synthesize loop invariants for programs with a single loop~\cite{code2inv18,ryan2020cln2inv} or multiple loops~\cite{de2016polynomial} on the numerical program. These approaches are unable to generate satisfiable and adequate specifications to fully prove the correctness of basic programs such as bubble sort.

Motivated by the research gap, {\name} is presented.
It synthesizes specification in a bottom-up manner, synthesizing versatile specifications (\ie, not only loop invariants, but also precondition, postcondition, and assigned variables, which are necessary for the full proof).
It validates the satisfiability of specifications whenever specifications are synthesized, and verifies the adequacy of specifications after all specifications are synthesized.

\begin{figure} [t]
    \centering
    \vspace{-5ex}
    \includegraphics[width=1.0\linewidth]{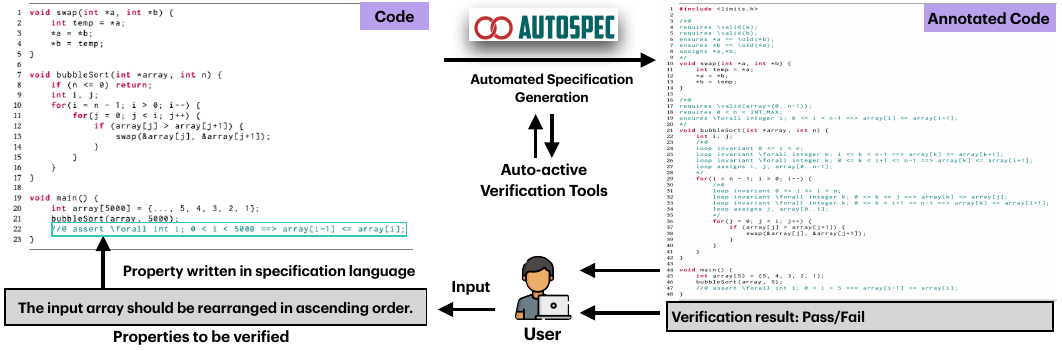}
    \setlength{\abovecaptionskip}{-10pt}
    \setlength{\belowcaptionskip}{-15pt}
    \caption{User Scenario of \name}
    \label{fig:pipeline}
\end{figure}

\textbf{\textit{User Scenario --}} 
We envision the user scenario of \name in 
Figure~\ref{fig:pipeline}.
Given a program and properties under verification, \name provides a fully automated verification process. It synthesizes the specifications for the program, validates the satisfiability of specifications, verifies the specifications against the desired properties, and outputs the verification result with proof if any. 

Note that proof can be provided by \name if the program is correctly implemented (\ie, the properties can be verified). When the given program is syntactically buggy, the program reports the syntactic error at the beginning before launching \name. If the given program is semantically buggy, then \name cannot synthesize adequate specifications for verification, the synthesis terminates when the maximum iteration number is reached. 

\section{Methodology}\label{sec:methd}
\begin{figure*}[t]
    \centering
    \vspace{-5ex}
    \includegraphics[width=1.0\linewidth]{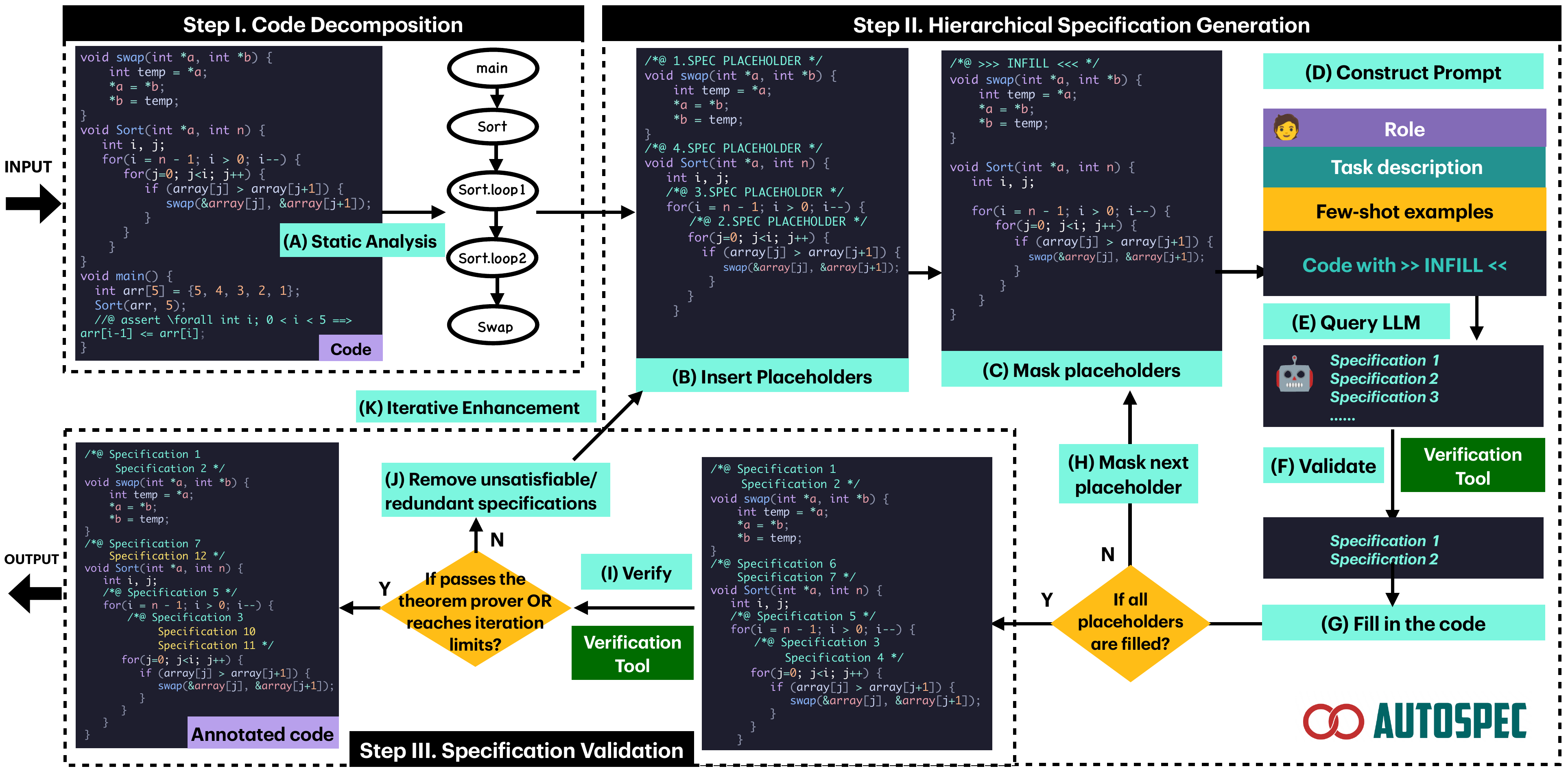}
    \setlength{\abovecaptionskip}{-5pt}
    \setlength{\belowcaptionskip}{-15pt}
    \caption{Overview of {\name}{\iffalse. \footnotesize{The input of \name is a C program with properties under verification, and the output is the annotated program with generated specifications, together with the verification result.}\fi}}
    \label{fig:overview}
\end{figure*}

Figure~\ref{fig:overview} shows an \textbf{overview} of {\name}. The workflow comprises three main steps: 
{\ding{202}} \textbf{Code Decomposition} (Section~\ref{sec:step1}). 
{\name} statically analyzes a \iffalse given\fi C program by decomposing it into a call graph, where loops are also represented as nodes. 
The \iffalse objective\fi aim of the first step is to generalize the procedure that was previously discussed in Section~\ref{sec:moti} to include the implicit knowledge of simulating interactions between humans and verification tools.
By decomposing the program into smaller components, LLMs can iteratively focus on different code components for a more comprehensive specification generation.
{\ding{203}} \textbf{Hierarchical specification generation} (Section~\ref{sec:step2}). 
Based on the call graph with loops, {\name} inserts \textit{placeholders} in each level of the graph in a bottom-up manner. Taking the program in Listing~\ref{fig:moti} for example, \name inserts the first placeholder (\texttt{\color{teal}/*@ 1. SPEC PLACEHOLDER */}) before \texttt{swap}, and then inserts the second placeholder in the inner loop of \texttt{Sort}. 
Then, \name iteratively masks the placeholder one at a time with ``\texttt{\color{teal}$>>>$ INFILL $<<<$}'' and feeds the masked code into LLMs together with few-shot examples. After querying LLMs, \iffalse LLMs\fi they reply with a set of specifications. {\name} then fills the generated specifications into the placeholder and proceeds to the next \iffalse placeholder\fi one. \iffalse Till\fi Once all the placeholders are filled with LLM-generated specifications, {\name} proceeds to the next step. 
{\ding{204}} \textbf{\textit{Specification Validation}} (Section~\ref{sec:step3}). 
\name feeds the verification conditions of each generated specification into a theorem prover to verify their satisfiability.
If the theorem prover confirms the satisfiability of the \iffalse generated\fi specifications, they will be annotated as a comment in the source code.
Otherwise, if the theorem prover identifies any unsatisfiable specifications \iffalse (\eg, one specification describes \texttt{\color{teal} ensures i < 0} but the actual value of i is always bigger than 0)\fi (\ie, cannot be satisfied by the program), \name removes those specifications and annotates program with the remaining specifications.
Then, {\name} returns to the second step to insert additional placeholders immediately after the specifications generated in the previous iteration and generate more specifications.
This iterative process continues until all the specifications are successfully verified by the {\iffalse theorem\fi} prover or until the number of iterations reaches the predefined upper limit (in our evaluation, this is set to 5). 
We will explain the methodology of each step in detail.

\subsection{Code Decomposition}\label{sec:step1}

Using static analysis, \name constructs a comprehensive call graph for the given program to identify the specific locations where specifications should be added and determine the order in which these specifications should be added. 
This call graph is an extended version of the traditional one, where loops are also treated as nodes, in addition to functions.
This is particularly useful for complex programs where loops can significantly affect the program’s behavior.
\iffalse Moreover, the extended graph helps \iffalse to\fi determine the locations where specifications should be added and the order in which to add specifications at these locations.\fi

The algorithm for constructing such a call graph is shown in Algorithm~\ref{alg:code_decomposition}. 
Specifically, the algorithm selects a function that contains the target assertion to be verified as the entry point for the call graph construction. 
\iffalse Then it starts by parsing the source code into an abstract syntax tree (AST). \fi
Then, it traverses the abstract syntax tree (AST) of the source code to identify all functions and loops \iffalse in the code\fi and their calling relationships.
For instance, the extended call graph \iffalse of\fi generated for the program in Listing~\ref{fig:moti} \iffalse can be seen\fi is given in Figure~\ref{fig:overview}(A).

Then, the specifications are generated step-by-step based on the nodes in the extended call graph.
When generating the specification for a node, one only needs to consider the code captured by the node and the specifications of its \iffalse adjacent nodes \fi callees in the extended call graph. Furthermore, 
modeling loops in addition to functions as separate nodes in the extended call graph allows \name to generate loop invariants, which are essential to program verification.
Therefore, code decomposition allows LLMs to focus on small \iffalse sections of the code\fi program components to generate specifications, thus reducing the complexity of specification generation and making it more manageable and efficient.
And traversing the extended call graph from bottom to top can simulate the programmers' verification process.

\iffalse
For instance, the extended call graph generated for the program in Listing~\ref{fig:moti} \iffalse can be seen\fi is given in Figure~\ref{fig:overview}(A).
The graph shows the optimal order for specification generation, starting from the function contract of the \texttt{swap} function, then the loop contract of the \texttt{bubbleSort} function, and finally the function contract of the \texttt{bubbleSort} function. This order echos to the process described in Section~\ref{sec:moti}.
\fi

\begin{figure}[t]
\vspace{-15pt}
\setlength{\textfloatsep}{0pt}% Remove \textfloatsep
\begin{algorithm}[H]
  \setstretch {0.90}
  \scriptsize
  \caption{Construct an Extended Loop/Call Graph}
  \label{alg:code_decomposition}
  \KwIn{The source code $C$ and the location $loc$ of the assertion to be verify}
  \KwOut{A call graph $G$ extended with loops}
  \BlankLine
    $G \leftarrow$ \texttt{get\_function}($loc$) ~\tcp*{\tiny initialize a Graph $G$}
    $WorkList$ $\leftarrow$ \texttt{get\_function}($loc$)\;
    \While{$WorkList != \emptyset$}{
        $Fn$ $\leftarrow$ \texttt{select\_and\_remove\_a\_node}($WorkList$) ~\tcp*{\tiny transitively visit all reachable nodes}
        \For{each basicBlock $bb$ in $Fn$}{ 
            \If(\tcp*[f]{\tiny If there exist a function call})
            {$bb$ calls a function $M$}{
                \If{$M$ is not already a node in $G$}{
                    Add a node $M$ to $G$\;
                    Add an edge from $Fn$ to $M$ in $G$\;
                    $WorkList$ $\leftarrow$ $WorkList$ $\cup$ $M$ ~\tcp*{\tiny Add this node (function) to the $WorkList$}
                }
            }
            \If(\tcp*[f]{\tiny If there exist a loop})
            {$bb$ is a loop entry for loop $L$}{
                \If{$L$ is not already a node in $G$}{
                    Add a node $L$ to $G$\;
                    Add an edge from $Fn$ to $L$ in $G$\;
                    $WorkList$ $\leftarrow$ $WorkList$ $\cup$ $L$ ~\tcp*{\tiny Add this node (loop) to the $WorkList$}
                }
            }
        }
    }
    \Return $G$ ~\label{algno:return_graph_end}
    % \vspace{-10pt}
\end{algorithm}
\vspace{-10pt}
\end{figure}

\subsection{Hierarchical Specification Generation}\label{sec:step2}

\name generates specifications for each node in the extended call graph in a hierarchical manner.
It starts from the leaf nodes and moves upward to the root node. 
This bottom-up approach ensures that the specifications for each function or loop are generated within the context of their callers.
Algorithm~\ref{alg:specification} shows the algorithm of hierarchical specification generation. 
The algorithm takes as input an extended call graph $G$, an iteration bound $t$, a large language model, and the assertion to be verified; and outputs an annotated code $C$ with generated specifications. 
In detail, the algorithm works as follows:
First, the algorithm initializes a code template $C$ with the original code without specifications (line~\ref{algno:initg}). 
The code template, similar to Figure~\ref{fig:overview}(B), includes placeholders. 
Each placeholder corresponds to a node in the call graph.
These placeholders will be iteratively replaced with the valid specifications generated by the LLMs and validated by the theorem prover, within a maximum of $t$ iterations (line~\ref{algno:iterate}).

In each iteration, the algorithm performs the following steps. 
\emph{First}, \name initializes a stack $S$ with the root node of graph $G$, which is the target function containing the assertion to be verified (line~\ref{algno:initstack}). \emph{Second}, \name pushes the nodes that require specification generation into the stack $S$ and traverses the stack $S$ in a depth-first manner (lines~\ref{algno:while_stack}-\ref{algno:while_end}). 
For each node $f$ in the stack, the algorithm checks if all the callees of $f$ have their specifications generated in this iteration (lines~\ref{algno:callee1}-\ref{algno:callee_gen}). If not, the algorithm pushes the callee nodes into the stack and marks $f$ as not ready for specification generation (lines~\ref{algno:push}-\ref{algno:allgen_false}). 
If all of the functions called by $f$ have had their specifications generated, the algorithm will then proceed to generate the specifications for $f$. (lines~\ref{algno:allgen_true}-\ref{algno:while_end}). 
In particular, \name queries the LLMs to generate a set of candidate specifications $spec_{tmp}$ for $f$ (line~\ref{algno:specgen}),  and validates $spec_{tmp}$ by examining their syntactic and semantic validity. 
Any illegal or unsatisfiable specifications are eliminated, and the remaining valid specifications are referred to as $spec_{f}$ (line~\ref{algno:validate}). 
The validation process may employ existing provers/verification tools to guarantee soundness. 
Then, \name inserts the validated specifications into the source code $C$ at the placeholder of $f$ (line~\ref{algno:insert}), and pops up the node $f$ from the stack, indicating that $f$ has its specification generated in this iteration (line~\ref{algno:while_end}). 
\emph{Third}, \iffalse upon traversing the stack $S$,\fi \name examines whether the whole verification task has been completed, \iffalse indicating\fi that is, whether \iffalse the annotated code $C$ with\fi the generated specifications \iffalse is sufficient \fi are adequate to verify the \iffalse given\fi target assertion (line~\ref{algno:assert}). 
If it does, \name proceeds to simplify the annotated code $C$ by eliminating redundant or unnecessary specifications (line~\ref{algno:simplify}) and then terminates (line~\ref{algno:break}). 
Otherwise, \iffalse we can logically\fi it is assumed that the specifications generated so far are satisfiable, though they may be inadequate. 
And \name will start another iteration to generate additional specifications while retaining \iffalse keeping the ones\fi those already generated.
\emph{Finally}, the algorithm returns the annotated code $C$ with the generated specifications as the output (line~\ref{algno:return_code_end}). 
After several iterations, if the whole verification task remains incomplete, the \iffalse developer\fi programmer can make a decision \iffalse regarding\fi on whether to involve professionals to continue with the verification process for the annotated code $C$.

\begin{figure}[t]
\vspace{-22.5pt}
\setlength{\textfloatsep}{0pt}% Remove \textfloatsep
\begin{algorithm}[H]
  \setstretch {0.90}
  \scriptsize
  \caption{Hierarchical Specification Generation}
  \label{alg:specification}
  \KwIn{A loop/call Graph $G$, an iteration bound $t$, a Large Language Model $LLM$, and the assertion $ass$ to be verified}
  \KwOut{Annotated Code $C$ with generated specifications}
  \BlankLine
    $C$.\texttt{init}() ~\tcp*{\tiny initialize the code (without specifications)} \label{algno:initg}
    \For(\tcp*[f]{\tiny iteratively enhancing specifications}){$i$ in \emph{\texttt{range}(0, $t$)}~\label{algno:iterate}}{
        Initialize a stack $S$ with the $root$ node of $G$~\label{algno:initstack}\;
        \While{$S$ is not empty~\label{algno:while_stack}}{
            $f$ = $S$.\texttt{top}() ~\tcp*{\tiny get the element at the top of the stack}
            $allgen$ = true\;
            \For{each $callee$ in $f$.\emph{\texttt{callees}()}~\label{algno:callee1}}{
                %\If{the specifications of $callee$ is not generated in this iteration~\label{algno:callee_gen}}{
                \If{ spec generation for $callee$ has not been done in $i^{th}$ iteration~\label{algno:callee_gen}}{
                    $S$.\texttt{push}($callee$) ~\label{algno:push}~\tcp*{\tiny push all callee into the stack} 
                    $allgen$ = false ~\label{algno:allgen_false}
                }
            }
            \If {$allgen$ == true~\label{algno:allgen_true}}{
                $spec_{t\!m\!p}$\,=\,\texttt{spec\_generation}($C$,\,$f$,\,$LLM$) ~\label{algno:specgen}~\tcp*{\tiny{\!query{\,}LLM{\,}to{\,}generate{\,}}specification{\,}candidates}
                $spec_f$ = \texttt{spec\_validation}($C$, $spec_{tmp}$) ~\label{algno:validate}~\tcp*{\tiny specification validation}
                $C$.\texttt{insert}($spec_f$) ~\label{algno:insert}~\tcp*{\tiny insert the specifications into the code template}
                $S$.\texttt{pop}() ~\label{algno:while_end}~\tcp*{\tiny pop up the top element $f$ of the stack} 
            }
        }
        \If(\tcp*[f]{\tiny determine whether the whole verification task has been completed}){\texttt{spec\_validation}($C$, $ass$)~\label{algno:assert}}{
            \texttt{simplify}($C$) ~\label{algno:simplify}~\tcp*{\tiny eliminating redundant specifications}
            \textbf{break} ~\label{algno:break}
        }
    }
    \Return $C$ ~\label{algno:return_code_end}
\end{algorithm}
\vspace{-15pt}
\end{figure}

Consider the extended call graph in Figure~\ref{fig:overview}(A) for example. \name first pushes the \texttt{main} function, the \texttt{sort} function, the \texttt{sort.loop1} loop, the \texttt{sort.loop2} loop, and the \texttt{swap} function into the stack in order, as all of them, except the \texttt{swap} function have some callee nodes that require specification generation. Then, it generates specifications for each function or loop in the stack in reserve order.
This order echoes what is described in Section~\ref{sec:moti}.
\name will leverage the power of LLMs to generate candidate specifications for each component/function (\ie, \texttt{spec\_generation} function in line~\ref{algno:specgen}).  
In the following, we discuss how \name utilizes LLMs for generating specifications.

\iffalse
Note that the process of human writing specifications inspires a hierarchical generation.
It leverages the power of LLMs to generate multiple candidate specifications for an independent component (\ie, \texttt{spec\_generation} function in line~\ref{algno:specgen}). That is, 
\iffalse This enables\fi the LLMs are able to focus on \iffalse to concentrate on one section of the code at a time\fi a (small) program component for specification generation, reducing the complexity of the task and improving the quality of the generated specifications.
In the following, we will discuss how we utilize LLMs for generating specifications.
\fi

\textbf{Specification generation by LLMs.} 
To employ LLMs in producing precise and reliable responses in the specified format, \name automatically generates a prompt for each specification generation task. This prompt is a natural language query that includes the role setting, task description, a few examples showing the desired specifications, 
and the source code with a highlighted placeholder (\eg, Figure~\ref{fig:overview}(C)). 
The prompt template used in \name is shown in Figure~\ref{fig:overview}(D).
Specifically, a prompt typically consists of the following elements: a system message, code with a placeholder, and an output indicator.
The system message provides the specification generation task description and the specification language, which are called \emph{context}. 
\name sets the \textbf{role} of LLMs as 
``\textit{As an experienced C/C++ programmer, I employ a behavioral interface specification language that utilizes Hoare style pre/post-conditions, as well as invariants, to annotate my C/C++ source code}''.
The system message also indicates the task's instructions, such as ``\textit{Fill in the \texttt{\color{teal}$>>>$ INFILL $<<<$}}''.
As explained in Section~\ref{sec:step1}, when querying LLMs for the specifications of a component (\ie, a function or loop), the code of this component and the specifications of its callees in the call graph are needed. That is to say, the irrelevant code that is not called by this component can be omitted, allowing LLMs to maintain their focus on the target component and reduces unnecessary token costs. Finally, \name uses the output format \texttt{\color{teal}/@ ... /} to indicate the generated specifications, which is crucial for programmatically processing the responses of LLMs.

To improve the quality of the generated specifications, \name employs the prompt engineering technique of few-shot prompting~\cite{nips20fewshot}.
To achieve this, the prompts are designed to include a few relevant input-output examples. 
Feeding LLMs a few examples can guide them in leveraging previous knowledge and experiences to generate the desired outputs. This, in turn, enables LLMs to effectively handle novel situations. 
In particular, few-shot prompting allows LLMs to facilitate the learning of syntax and semantics of specification language through in-context learning.
For example, consider a prompt that includes an input-output example with a loop invariant for an array that initializes all elements to 0, such as {\texttt{\color{teal} $\backslash$forall integer j; 0 <= j < i ==> ((char*)p)[j] == 0;}}.
With this example, \name is able to generate a valid loop invariant that involves using quantifiers for the inner loop of \texttt{bubbleSort} in a single query.

\subsection{Specification Validation}\label{sec:step3}
The hierarchical specification generation algorithm also employs specification validation (\ie, \texttt{spec\_validation()} in line~\ref{algno:validate}) and specification simplification (\ie, \texttt{simplify()} in line~\ref{algno:simplify}) techniques to ensure the quality \iffalse and efficiency\fi of the specifications.

{\textbf{Specification validation.}}
Once candidate specifications have been generated for a component, \name will check their syntactic and semantic validity (\ie, legality and satisfiability), as shown in Figure~\ref{fig:overview}(F).
Specifically, for a function, the legality and satisfiability of the generated specifications are checked immediately.
While for a loop, the legality is checked immediately, but the satisfiability check is postponed until the outermost loop. This is because inner loops often use variables defined in some of their outer loops (\eg, variable \texttt{i} in the \texttt{bubbleSort} example), and the satisfiability of all loop invariants needs to be verified simultaneously.

\name leverages the verification tool (\ie, \framac) to verify the specifications.
If the verification tool returns a compilation error, \name identifies the illegal specification where the error occurs and continues verifying without \iffalse eliminates\fi it if there are still some candidates. 
Otherwise, if the verification tool returns a verification failure,  \name identifies the unsatisfiable specification which fails during verification and continues verifying without \iffalse eliminates\fi it if there are still some candidates.
Finally, if the verification succeeds, the specifications will be correspondingly inserted into the code as a comment (Figure~\ref{fig:overview}(G)).

In addition, \name also validates whether the generated specifications are adequate to verify the target assertion (line~\ref{algno:assert}), which is the same as the validation above but with the target assertion.
Note that, the validation phase is crucial to \name as it ensures that the generated specifications are not only legal and satisfiable but also adequate to verify the target assertion (Figure~\ref{fig:overview}(I)).

{\textbf{Specification simplification} (Optional).}
The objective of specification simplification is to provide users with a concise and elegant specification that facilitates manual inspection and aids in understanding the implementation.
This process could be \emph{optional} if one's goal is simply to complete the verification task without placing importance on the specifications.
After successfully verifying the assertion, we proceed to systematically remove specifications that are not needed for their verification, one by one. 
Our main idea is that a specification is unnecessary if the assertion is still verifiable without it. We repeat this process until we reach the minimal set of specifications for manual reading.

\iffalse The specification simplification process eliminates some specifications for two reasons:\fi
There are two main reasons to eliminate specifications: 
(1) The specification is considered weak and does not capture relevant properties of the verification task. For example, both the loop invariant \texttt{\color{teal}i > 0} and \texttt{\color{teal}i > 1} are satisfiable, but \texttt{\color{teal}i > 0} can be safely removed. 
(2) The specification is semantically similar to another specification. 
For example, both \texttt{\color{teal}$\backslash$forall integer i; 0 < i < n ==> array[i-1] <= array[i];} and \texttt{\color{teal}$\backslash$forall integer i; 0 <= i < n-1 ==> array[i] <= array[i+1];} accurately describe the post-condition of \texttt{bubbleSort}.  Removing either of them has no impact on the verification results.

\section{Evaluation}\label{sec:eval}
% We have implemented {\name} and applied it to the C software analysis platform {\framac} and its deductive verification plugin {\wpplugin}. 
%This section presents a comprehensive evaluation of the effectiveness and usefulness of {\name}. 

\begin{table*}[t]
\centering
\setlength{\abovecaptionskip}{-0pt}
\setlength{\belowcaptionskip}{-5pt}
\caption{Statistics of Benchmarks.}
\label{tab:benchmark}
\renewcommand\arraystretch{1.0}
    \resizebox{1.025\linewidth}{!}{
\begin{tabular}{l|l|c|l|c|c}
\toprule
\textbf{Benchmarks\,/\,Project} & {\textbf{Description}} & {\footnotesize\begin{tabular}[c]{@{}c@{}}\textbf{Num\,of}\\\textbf{Prog}\end{tabular}} & {\normalsize \textbf{Types of Specifications}} & {\footnotesize\begin{tabular}[c]{@{}c@{}}\textbf{Ave}\\\textbf{LoC}\end{tabular}} & {\footnotesize\begin{tabular}[c]{@{}c@{}}\textbf{Num\,of}\\\textbf{Spec}\end{tabular}} \\
\midrule
\textbf{{\framac}-problems}\cite{frama-c-problems} & Programs\,with\,function\,calls,\,nested/multiple\,loops,\,arrays,\,pointers.

& 51 & pre/post-conditions,\,loop invariants & 17.43 & 1$\sim$3 \\
\textbf{X509-parser}~\cite{x509-parser} & A real-world software implements a X.509 certificate parser. & 6 & pre/post-conditions,\,loop invariants & 82.33 & 3$\sim$19 \\
\textbf{SyGuS}~\cite{SyGuS} & Programs with a single loop. & {133} & loop invariants & 22.56 & 1$\sim$12 \\
\textbf{OOPSLA-13}~\cite{oopsla13} & Programs with a single loop or nested/multiple loops. & 46 & loop invariants & 30.28 & 1$\sim$3 \\
\textbf{SV-COMP}~\cite{SVCOMP} & Programs with more complex nested/multiple loops. & 21 & loop invariants & 24.33 & 1$\sim$5\\
\bottomrule
\end{tabular}
}
\vspace{-3.5ex}
\end{table*}
\vspace{-0.5ex}
The {\iffalse conducted\fi} experiments aim to answer the following research questions:

\begin{steps}\setlength{\itemsep}{-0.5cm}
    \vspace{-1.5ex}
    \item {\textbf{Can \name generate specifications for various properties effectively?} \small{We aim to comprehensively characterize the effectiveness of \name against various types of specifications including pre/post-conditions, loop invariants.}} \\
    \vspace{1ex}
    \item {\textbf{Can \name generate specifications for loop invariant effectively?} \small{Loop invariant, as a major specification type, is known for its difficulty and significance. We select three benchmarks with linear and nested loop structures and compare them with state-of-the-art approaches.}}
    \\
    \vspace{1ex}
    \item {\textbf{Is \name efficient?} \small{We compare the \name's overhead incurred by LLM querying and theorem proving with the baselines\iffalse, showing the efficiency of \name\fi.}}
    \\
    \vspace{1ex}
    \item {\textbf{Does every step of \name contribute to the final effectiveness?} \small{We conduct an ablation study on each part of the \name's design, showing the distinct contribution made independently.}}
    \vspace{-1.5ex}
\end{steps}

% \usepackage{multirow}
\begin{table*}[t!]
\centering
\setlength{\abovecaptionskip}{-0pt}
\setlength{\belowcaptionskip}{-25pt}
\caption{Effectiveness of \name in General Specification Generation\iffalse . {\footnotesize The benchmark \framac-problem contains 51 C programs with various types of specifications.}\fi}
\label{tab:rq1}
\scriptsize
\renewcommand\arraystretch{1.05}
    \resizebox{1.0\linewidth}{!}{
\begin{tabular}{l|lrc|cllccr}
\toprule
\multicolumn{4}{c|}{\textbf{Benchmark Information}} & \multicolumn{6}{c}{\cellcolor[HTML]{EBCB8B}\textbf{\name}} \\ \midrule
\textbf{Type} & \textbf{Program} & \textbf{LoC} & \begin{tabular}[c]{@{}l@{}}\textbf{Component}\\ (\textbf{Func}, \textbf{Loop})\end{tabular} & \textbf{Success} & \textbf{Ratio} & \textbf{ Iterations} & \begin{tabular}[c]{@{}c@{}}\textbf{Generated}\\ \textbf{Spec}\end{tabular} & \begin{tabular}[c]{@{}c@{}}\textbf{Correct}\\ \textbf{Spec}\end{tabular} & \begin{tabular}[c]{@{}c@{}}\textbf{Time(s)}\\ \textbf{mean $\pm$ std}\end{tabular} \\ \hline
\multirow{12}{*}{general\_wp\_problems} & absolute\_value.c & 15 & 1 (1,0) & \ding{52} & 5/5 & 1,1,1,1,1 & 7,7,7,7,7 & 7/7 &     $ 14.17 \pm 8.74 $  \\
 & add.c & 11 & 1 (1,0) & \ding{52} & 5/5 & 1,1,1,1,1 & 3,3,3,3,3 & 2/2 &    $ 14.36 \pm 8.20 $   \\
 & ani.c & 18 & 2 (1,1) & \ding{53} & 0/5 & -,-,-,-,- & 41,33,26,20,24 & 3/4 &   --  \\
 & diff.c & 10 & 1 (1,0) & \ding{52} & 5/5 & 1,1,1,1,1 & 2,2,2,2,2 & 1/1 &   $ 8.85 \pm 5.72 $   \\
 & gcd.c & 22 & 1 (1,0) & \ding{53} & 0/5 & -,-,-,-,- & 3,6,8,5,5 & 2/5 &    --  \\
 & max\_of\_2.c & 15 & 1 (1,0) & \ding{52} & 5/5 & 1,1,1,1,1 & 4,3,5,3,5 & 2/2 &     $ 17.64 \pm 11.24 $ \\
 & power.c & 18 & 2 (1,1) & N/A & -- & -- & -- & -- &     --  \\
 & simple\_interest.c & 14 & 1 (1,0) & \ding{52} & 5/5 & 1,1,1,1,1 & 5,5,5,5,5 & 5/5 &   $ 15.34 \pm 8.60 $   \\
 & swap.c & 16 & 1 (1,0) & \ding{52} & 5/5 & 1,1,1,1,1 & 3,3,3,3,3 & 2/2 &   $ 15.36 \pm 8.66 $  \\
 & triangle\_angles.c & 14 & 1 (1,0) & \ding{52} & 5/5 & 1,1,1,1,1 & 7,5,6,4,5 & 4/4 &   $ 23.99 \pm 13.67 $ \\
 & triangle\_sides.c & 16 & 1 (1,0) & \ding{52} & 5/5 & 1,1,1,1,1 & 3,3,3,2,2 & 2/2 &    $ 20.11 \pm 11.41 $ \\
 & wp1.c & 14 & 1 (1,0) & \ding{53} & 0/5 & -,-,-,-,- & 1,4,1,4,2 & 3/3 &    --  \\ \hline
\multirow{8}{*}{pointers} & add\_pointers.c & 19 & 1 (1,0) & \ding{52} & 5/5 & 1,1,1,1,1 & 3,4,4,4,4 & 2/2 &     $ 10.00 \pm 5.85 $ \\
 & add\_pointers\_3\_vars.c & 20 & 1 (1,0) & \ding{52} & 5/5 & 2,5,3,-,3 & 3,6,10,14,4 & 3/3 &   $ 74.45 \pm 47.78 $ \\
 & div\_rem.c & 12 & 1 (1,0) & \ding{52} & 5/5 & 1,1,1,1,1 & 7,8,7,7,7 & 4/4 &   $ 14.66 \pm 8.39 $  \\
 & incr\_a\_by\_b.c & 13 & 1 (1,0) & \ding{52} & 5/5 & 1,1,1,1,1 & 6,6,6,4,6 & 3/3 &     $ 18.82 \pm 13.43 $ \\
 & max\_pointers.c & 16 & 1 (1,0) & \ding{52} & 5/5 & 1,2,1,1,1 & 5,4,4,5,5 & 4/4 &  $ 41.73 \pm 15.74 $ \\
 & order\_3.c & 36 & 1 (1,0) & \ding{53} & 0/5 & -,-,-,-,- & 16,19,15,6,12 & 3/4 &   --  \\
 & reset\_1st.c & 16 & 1 (1,0) & \ding{52} & 5/5 & 1,1,1,1,1 & 7,5,4,6,5 & 4/4 &     $ 15.23 \pm 9.42 $  \\
 & swap\_pointer.c & 13 & 1 (1,0) & \ding{52} & 5/5 & 1,1,1,1,1 & 5,5,5,5,5 & 2/2 &  $ 10.72 \pm 6.00 $ \\ \hline
\multirow{8}{*}{loops} & 1.c & 9 & 1 (0,1) & \ding{52} & 5/5 & 1,1,1,1,1 & 2,2,2,2,2 & 1/1 &     $ 2.35 \pm 2.23 $   \\
 & 2.c & 17 & 2 (1,1) & \ding{53} & 0/5 & -,-,-,-,- & 8,10,11,11,5 & 2/5 &  --  \\
 & 3.c & 18 & 2 (1,1) & \ding{53} & 0/5 & -,-,-,-,- & 10,6,7,5,6 & 3/4 &     --  \\
 & 4.c & 18 & 2 (1,1) & N/A & -- & -- & -- & -- &     --  \\
 & fact.c & 19 & 2 (1,1) & \ding{53} & 0/5 & -,-,-,-,- & 3,8,7,6,6 & 3/7 &   --  \\
 & mult.c & 16 & 2 (1,1) & \ding{53} & 0/5 & -,-,-,-,- & 6,14,10,9,16 & 2/3 &   --  \\
 & sum\_digits.c & 17 & 2 (1,1) & \ding{53} & 0/5 & -,-,-,-,- & 7,13,13,12,11 & - &  --  \\
 & sum\_even.c & 16 & 2 (1,1) & \ding{53} & 0/5 & -,-,-,-,- & 9,14,16,18,19 & 2/3 &  --  \\ \hline
\multirow{8}{*}{immutable\_arrays} & array\_sum.c & 16 & 2 (1,1) & \ding{53} & 0/5 & -,-,-,-,- & 5,4,4,3,4 & 3/5 &   --  \\
 & binary\_search.c & 24 & 2 (1,1) & \ding{52} & 1/5 & 3,-,-,-,- & 16,16,30,36,19 & 7/7 &    $ 739.62 \pm 239.59 $   \\
 & check\_evens\_in\_array.c & 19 & 2 (1,1) & \ding{53} & 0/5 & -,-,-,-,- & 11,18,15,13,16 & 4/6 &   --  \\
 & max.c & 20 & 2 (1,1) & \ding{52} & 5/5 & 1,1,1,1,1 & 10,9,12,8,7 & 5/5 &  $ 40.11 \pm 29.49 $ \\
 & occurences\_of\_x.c & 26 & 2 (1,1) & \ding{52} & 5/5 & 2,1,1,1,2 & 16,12,10,14,12 & 3/3 &     $ 121.83 \pm 75.04 $    \\
 & sample.c & 19 & 1 (0,1) & \ding{52} & 5/5 & 1,1,1,1,1 & 3,3,4,3,3 & 1/1 &     $ 16.89 \pm 10.47 $ \\
 & search.c & 17 & 2 (1,1) & \ding{53} & 0/5 & -,-,-,-,- & 12,14,16,16,12 & 5/8 &    --  \\
 & search\_2.c & 18 & 2 (1,1) & \ding{52} & 4/5 & 1,3,2,-,3 & 13,18,19,10,16 & 5/5 &     $ 155.32 \pm 175.44 $   \\ \hline
\multirow{2}{*}{mutable\_arrays} & array\_double.c & 19 & 2 (1,1) & \ding{52} & 4/5 & -,2,2,2,3 & 12,17,19,16,17 & 4/4 &  $ 81.44 \pm $ 21.14 \\
 & bubble\_sort.c & 26 & 3 (1,2) & \ding{52} & 3/5 & -,2,3,3,- & 9,12,15,12,15 & 10/10 &     $ 448.76 \pm 554.81 $   \\ \hline
\multirow{3}{*}{more\_arrays} & equal\_arrays.c & 15 & 2 (1,1) & \ding{53} & 0/5 & -,-,-,-,- & 9,14,15,13,8 & 5/7 &     --  \\
 & replace\_evens.c & 17 & 2 (1,1) & \ding{52} & 5/5 & 1,1,1,1,1 & 13,12,15,20,14 & 3/3 &    $ 52.33 \pm 17.30 $  \\
 & reverse\_array.c & 23 & 2 (1,1) & \ding{53} & 0/5 & -,-,-,-,- & 10,7,14,13,18 & 5/- &    --  \\ \hline
\multirow{5}{*}{arrays\_and\_loops} & 1.c & 10 & 1 (1,0) & \ding{52} & 5/5 & 1,1,1,1,1 & 3,2,3,2,3 & 1/1 &   $ 3.22 \pm 2.07 $   \\
 & 2.c & 18 & 2 (1,1) & \ding{52} & 5/5 & 1,1,1,1,1 & 10,10,10,10,10 & 2/2 &     $ 30.15 \pm 27.79 $ \\
 & 3.c & 19 & 1 (1,0) & \ding{52} & 5/5 & 1,1,1,1,1 & 9,10,10,4,9 & 2/2 &    $ 12.05 \pm 7.09 $  \\
 & 4.c & 18 & 2 (1,1) & \ding{52} & 5/5 & 1,1,1,1,1 & 12,10,10,8,0 & 2/2 &   $ 18.34 \pm 13.46 $ \\
 & 5.c & 18 & 2 (1,1) & \ding{53} & 0/5 & -,-,-,-,- & 12,10,10,4,9 & 3/4 &   --  \\ \hline
\multirow{5}{*}{miscellaneous} & array\_find.c & 20 & 2 (1,1) & \ding{53} & 0/5 & -,-,-,-,- & 7,7,7,7,7 & 4/7 &  --  \\
 & array\_max\_advanced.c & 20 & 2 (1,1) & \ding{52} & 5/5 & 1,1,1,1,1 & 5,6,5,6,5 & 2/2 &   $ 31.99 \pm 34.41 $ \\
 & array\_swap.c & 18 & 1 (1,0) & \ding{52} & 5/5 & 1,1,1,1,1 & 8,4,7,8,4 & 3/3 &    $ 26.05 \pm 30.72 $ \\
 & increment\_arr.c & 17 & 2 (1,1) & \ding{53} & 0/5 & -,-,-,-,- & 2,2,2,2,2 & 3/6 &     --  \\
 & max\_of\_2.c & 14 & 1 (1,0) & \ding{52} & 5/5 & 1,1,1,1,1 & 2,3,3,2,3 & 1/1 &     $ 9.98 \pm 10.31 $  \\
 \midrule 
 \textbf{Overall} & & & &  \textbf{31 / 51} & & & & & $ 89.17 \pm 172.75 $\\
 \bottomrule
 
\end{tabular}
}
\vspace{-5pt}
\end{table*}

\subsection{Evaluation Setup}

\subsubsection{\textbf{Benchmark}.}
{\iffalse To assess the effectiveness of {\name} in generating specifications, we\fi} We conducted evaluations on four benchmarks and a real-world project. The statistical details of these benchmarks can be found in Table~\ref{tab:benchmark}.
The {\framac}-problems~\cite{frama-c-problems} benchmark and the X509-parser~\cite{x509-parser} comprises programs that involve multiple functions or loops, requiring the formulation of pre/post-conditions, loop invariants, \etc.
The SyGuS~\cite{SyGuS} benchmark only includes programs with linear loop structures.
While the OOPSLA-13~\cite{oopsla13} and SV-COMP~\cite{SVCOMP} benchmarks include programs with nested or multiple loops, making them suitable for evaluating the versatility and diversity of generated specifications. 
{Please note that we assume the programs being verified are free of compilation errors, and the properties being verified are consistent with the programs.
If there are any inconsistencies between the code and properties, \name is expected to fail the verification after the iterations end.}

\vspace{-3ex}

\subsubsection{\textbf{Baselines}.} For RQ1, as previous works have primarily relied on manually written specifications for the deductive verification of functional correctness for C/C++ programs~\cite{baudin2020wp,blanchard2019towards}, we then \iffalse only\fi conduct our approach based on this baseline, and use the ablation study to \iffalse show\fi demonstrate the contribution of different parts of the design in {\name} in RQ4. 
For RQ2, we compare with Code2Inv~\cite{code2inv18}, a learning-based approach for generating linear loop invariants\footnote{\iffalse For every C program, it instruments the program, collects the change of variable values during execution, and trains a deep neural network using reinforcement learning.\fi We reproduce their implementation using the provided replicable package and run the tool on two additional benchmarks following their instructions. However, in their original setting, the maximal time limit for each program is set to 12 hours, which is far from affordable. So we lowered the threshold to 1 hour for efficiency.}. Although there are newer approaches built on \iffalse top of\fi Code2Inv such as CLN2INV~\cite{ryan2020cln2inv}, their replicable toolkit is only applicable \iffalse in\fi to the benchmark they used (\ie, SyGuS~\cite{SyGuS}) and incomplete, failing to apply to other benchmarks. Additionally, for RQ2, we also compared with Pilat~\cite{de2016polynomial} \iffalse under\fi using the default settings. 

\vspace{-3ex}

\subsubsection{\textbf{Configuration}.}
For implementation, we use ChatGPT's API \textit{gpt-3.5-turbo-0613}. 
We configure the parameters in API as follows: \texttt{max\_token}\iffalse (the length of the output in tokens)\fi: $2048$, \texttt{temperature}\iffalse (the randomness of the generated text, the lower the stabler)\fi: $0.7$. 
To show the generalizability of \name, we also utilize \textit{Llama-2-70b} for conducting a comparable experiment (Section~\ref{sec:threat}). Lastly, we employ {\framac}~\cite{frama-c,kirchner2015frama} and its {\wpplugin} plugin to verify the specifications.
{\iffalse\xzw{iteration bound t?}\wc{omit, mentioned it in method}\fi}

%%%%%%%%%%%%%%%%%%%%%%%%%%%%%%%%%%%%%%%%%%%%%%%%%%
%%%%%%%%%%%%%%%%%%%%%%%% RQ1 %%%%%%%%%%%%%%%%%%%%%
%%%%%%%%%%%%%%%%%%%%%%%%%%%%%%%%%%%%%%%%%%%%%%%%%%
\subsection{RQ1. Effectiveness on General Specification} \label{sec:rq1}
Table~\ref{tab:rq1} shows the results of \name on the \framac-problems{\iffalse\xzw{problems? better to use ``problem''}\wc{the benchmark's name use "problems"}\fi} benchmark. This benchmark consists of 51 C programs, {\iffalse~\footnote{Two programs in the \framac-problems benchmark are incompatible, \ie, \texttt{power.c} under \texttt{general\_wp\_problems} and \texttt{4.c} under \texttt{loops}. See Section~\ref{sec:case-study} for details.\fi} 
divided into eight categories (as indicated in the entry \textit{Type}){\iffalse, including \texttt{general\_wp\_problem} (\ie, programs with weakest preconditions), \texttt{pointers} (\ie, programs with pointer usages), \etc\fi}. Each type contains several programs. The size of the programs ranges from 9 to 36 lines of code (the entry \textit{LoC}). We also list the number of functions and loop structures defined in the program. Most programs contain a main function, with one or more loop structures. Since we could not find other previous work that can automatically generate various types of specifications to complete the verification task on \framac-problems benchmark, we hereby show the effectiveness of \name in detail. 

Overall, 31/51 of these programs can be successfully solved by \name.
In particular, due to the randomness of LLMs, we ran the experiment five times for each program and reported the detailed results. The success rate is tabulated in Table~\ref{tab:rq1}, column \textit{Ratio}. It shows that the results are stable over five runs. Almost all passed cases can be successfully solved in five runs, with only a few exceptions (\eg,\,1/5,\,4/5). The stable result shows that the randomness of LLMs has little impact on the effectiveness of \name. 
\iffalse In addition,\fi Furthermore, \name enables an iterative enhancement on specification generation. We hereby show the number of iterations \iffalse that are\fi used for success generation (column \textit{Iterations}). Most cases can be solved in the first iteration. While the iterative enhancement also contributes to certain improvements. For example, \texttt{add\_pointers\_3\_vars.c} in the \texttt{pointers} category needs two more iterations to generate adequate specifications to pass the theorem prover. \iffalse Furthermore\fi In addition, we also report the number of generated specifications that are correct by using the ground truth in the benchmark as a reference, as shown in column \textit{Correct Spec}. We can see that for the failed cases, there is at least one generated specification that is correct. This shows that the generated specifications are not excessive, and still have the potential to improve.
Finally, in terms of overhead (column \textit{Time(s)}), \name processes \iffalse one\fi a case in minutes, from 2.53 seconds to 12 minutes, with an average of 89.17 seconds. 

% \usepackage{multirow}
\begin{table*}[t!]
\centering
\setlength{\abovecaptionskip}{-0pt}
\setlength{\belowcaptionskip}{-25pt}
\caption{Effectiveness on a Real-world X.509 Certificate Parser Project}
\label{tab:rq1-real-world}
\scriptsize
\renewcommand\arraystretch{1.15}
    \resizebox{1.0\linewidth}{!}{
\begin{tabular}{l|llr|cllccr}
\toprule
\multicolumn{4}{c|}{\textbf{Function Information}} & \multicolumn{6}{c}{\cellcolor[HTML]{EBCB8B}\textbf{\name}} \\
\midrule
\textbf{Project} & \textbf{Function} & \textbf{Feature} & \textbf{LoC} & \textbf{Success} & \textbf{Ratio} & \textbf{ Iterations} & \begin{tabular}[c]{@{}c@{}}\textbf{Generated}\\ \textbf{Spec}\end{tabular} & \begin{tabular}[c]{@{}c@{}}\textbf{Correct}\\ \textbf{Spec}\end{tabular} & \begin{tabular}[c]{@{}c@{}}\textbf{Time(s)}\\ \textbf{mean $\pm$ std}\end{tabular} \\ 
\midrule
 \rowcolor{rowcolor} \cellcolor{white}\multirow{9}{*}{X509-parser} &\texttt{check\_ia5\_string} & loop; buffer pointer & 60 & \ding{52} & 5/5 & 1,1,2,1,1 & 13,11,14,13,11 & 6/6 & $ 20.89 \pm 10.51 $ \\
 & \texttt{verify\_correct\_time\_use} & switch-case  & 90 & \ding{52} & 5/5 & 1,1,3,1,2 & 10,19,23,15,16 & 3/3 & $ 24.16 \pm 11.73 $ \\
 \rowcolor{rowcolor} \cellcolor{white} & \texttt{bufs\_differ} & loop; buffer pointer  & 55 & \ding{52} & 5/5 & 1,1,1,1,1 & 17,17,20,16,15 & 5/5 & $ 12.49 \pm 5.11 $ \\
 & \texttt{parse\_null} & \begin{tabular}[l]{@{}l@{}}call the function bufs\_differ;\\buffer pointer\end{tabular}  & 87 & \ding{52} & 2/5 & -,-,1,1,- & 25,36,44,33,34 & 13/13 & $ 260.96 \pm 118.58 $ \\
 \rowcolor{rowcolor} \cellcolor{white} & \texttt{parse\_algoid\_params\_none} & \begin{tabular}[l]{@{}l@{}}call the function \texttt{parse\_null}\\and \texttt{bufs\_differ}\end{tabular}  & 136 & \ding{52} & 2/5 & -,2,-,-,2 & 184,92,156,142,96 & 19/19 & $ 957.14 \pm 446.56 $ \\
 % \texttt{time\_components\_to\_comparable\_u64}
 & \texttt{time\_components} & \begin{tabular}[l]{@{}l@{}}shift operation; \\multiple data type\end{tabular} & 63 & \ding{52} & 5/5 & 1,1,1,1,1 & 11,17,16,13,17 & 7/7 & $ 11.82 \pm 5.34 $ \\
\midrule
Overall &  &  & &  \textbf{6 / 6} & & & & & $214.58 \pm 389.58 $ \\
\bottomrule
 
\end{tabular}
}
\vspace{-5pt}
\end{table*}

\textbf{A real-world X509 parser project.}
The X509-parser project, which aims to ensure the absence of runtime errors, has undergone verification by \framac and the ACSL specification language.
Note that the specifications for this project were manually added throughout 5 months\,\cite{ebalard2019journey}.
It is currently impractical to seamlessly apply \name to the entire project without human intervention. 
We manually extracted 6 representative functions without specifications.
These functions handle pointer dereference, multiple data types, shift operations, \etc.
For each function, we set a verification target that accurately describes its functional correctness properties.
\name generates specifications for these functions, as shown in Table~\ref{tab:rq1-real-world}.
Surprisingly, all 6 functions were solved by \name.
Through our comprehensive manual examination of the generated specifications, we found that \name can generate a variety of specifications {\iffalse that were\fi}not previously \iffalse authored\fi written by the developer.
These specifications play a crucial role in ensuring functional correctness.
Considering that it takes five calendar months \iffalse for writing\fi to write specifications for the whole X509-parser project~\cite{ebalard2019journey}, \name can automatically generate the required specifications for the functions in X509-parser in a few minutes.
We believe that \name could be useful for real-world verification tasks.

\begin{table*}[t!]
\centering
\setlength{\abovecaptionskip}{-0pt}
\setlength{\belowcaptionskip}{-30pt}
\caption{Effectiveness on Loop Invariants Synthesis{\iffalse . \footnotesize SyGuS~\cite{SyGuS} benchmark contains 133 C programs with one loop structure. While OOPSLA-13~\cite{oopsla13} benchmark contains 46 C programs with one loop, multiple loops, and nested loop structures. This table elaborates partial results on three baselines (Pailt, Code2Inv, and CLN2INV) compared with \name in the number of successfully verified programs and the overhead in seconds.\fi}}
\label{tab:rq2-merge}
\small
\renewcommand\arraystretch{1.2}
\resizebox{1.0\linewidth}{!}{
\begin{tabular}{ll|crcccrcr||rrrr|crcrrc}
\toprule
\multicolumn{10}{c||}{\normalsize\cellcolor[HTML]{CBE0B8}\textbf{SyGuS~\cite{SyGuS} (133 C Programs with One Loop)}} & \multicolumn{10}{c}{\normalsize\cellcolor[HTML]{D2E1F1}\textbf{OOPSLA-13~\cite{oopsla13} (46 C Programs with Various Loop Types)}} \\
\midrule
\multicolumn{2}{c|}{\textbf{Info}} & \multicolumn{2}{c}{\cellcolor[HTML]{EBCB8B}\textbf{\name}} & \multicolumn{2}{c}{\textbf{Pailt}} & \multicolumn{2}{c}{\textbf{Code2Inv}} & \multicolumn{2}{c||}{\textbf{CLN2Inv}} & \multicolumn{4}{c|}{\textbf{Info}} & \multicolumn{2}{c}{\cellcolor[HTML]{EBCB8B}\textbf{\name}} & \multicolumn{2}{c}{\textbf{Code2Inv}} & \multicolumn{2}{c}{\textbf{CLN2INV}}\\
\midrule
ID & LoC & Sucess & Time(s) & Success & Time(s) & Success & Time(s) & Success & Time & ID & LoC & Type & \begin{tabular}[c]{@{}l@{}}Loop\\ Num.\end{tabular} & Success & \begin{tabular}[c]{@{}c@{}}\textbf{Time(s)}\\ \textbf{mean $\pm$ std}\end{tabular} & Success & Time(s) & Success & Time(s)\\
\rowcolor{rowcolor}
1 & 29 & \ding{52} & 6.65 & \ding{53} & -- & \ding{52} & 4950.78 & \ding{52} & 4.32 & 1 & 23 & Linear & 1 & \ding{52} & $ 6.25\pm7.22 $ & \ding{52} & 337.8 & \ding{108} & --\\
2 & 20 & \ding{52} & 5.97 & \ding{53} & -- & \ding{53} & -- & \ding{52} & 4.11 & 2 & 27 & Linear & 1 & \ding{52} & $ 7.17\pm7.38 $ & \ding{52} & 74.36 & \ding{108} & --\\
\rowcolor{rowcolor}
3 & 18 & \ding{53} & 36.24 & \ding{53} & -- & \ding{53} & -- & \ding{52} & 0.22 & 3 & 22 & Linear & 1 & \ding{52} & $ 113.47\pm88.10 $ & \ding{52} & 46.22 & \ding{53} & --\\
4 & 13 & \ding{53} & 32.28 & \ding{53} & -- & \ding{53} & -- & \ding{52} & 0.21 & 4 & 28 & Linear & 1 & \ding{52} & $ 6.04\pm7.64 $ & \ding{53} & -- & \ding{108} & --\\
\rowcolor{rowcolor}
5 & 17 & \ding{52} & 19.22 & \ding{53} & -- & \ding{53} & -- & \ding{52} & 2.48 & 5 & 30 & Linear & 1 & \ding{52} & $ 8.59\pm9.80 $ & \ding{53} & -- & \ding{108} & --\\
6 & 21 & \ding{53} & 142.95 & \ding{53} & -- & \ding{53} & -- & \ding{52} & 1.7 & 6 & 31 & Linear & 1 & \ding{52} & $ 41.62\pm24.06 $ & \ding{53} & -- & \ding{53} & -- \\
\rowcolor{rowcolor}
7 & 17 & \ding{52} & 57.15 & \ding{53} & -- & \ding{52} & 128.03 & \ding{52} & 3.47 & 7 & 30 & Linear & 1 & \ding{52} & $ 12.66\pm15.12 $ & \ding{53} & -- & \ding{108} & --\\
8 & 15 & \ding{52} & 92.75 & \ding{53} & -- & \ding{52} & 72.74 & \ding{52} & 3.13 & 8 & 24 & Linear & 1 & \ding{52} & $ 3.40\pm3.25 $ & \ding{53} & -- & \ding{108} & --\\
\rowcolor{rowcolor}
9 & 25 & \ding{52} & 39.66 & \ding{53} & -- & \ding{53} & -- & \ding{52} & 3.04 & 9 & 27 & Linear & 1 & \ding{53} & -- & \ding{53} & -- & \ding{53} & --\\
10 & 21 & \ding{53} & 295.83 & \ding{53} & -- & \ding{52} & 53.39 & \ding{52} & 3.14 & 10 & 26 & Linear & 1 & \ding{52} & $ 9.40\pm7.29 $ & \ding{53} & -- & \ding{108} & --\\
\rowcolor{rowcolor}
11 & 18 & \ding{52} & 77.48 & \ding{53} & -- & \ding{52} & 145.02 & \ding{52} & 3.33 & 11 & 26 & Linear & 1 & \ding{52} & $ 16.82\pm16.90 $ & \ding{53} & -- & \ding{53} & --\\
12 & 30 & \ding{52} & 106.137 & \ding{53} & -- & \ding{52} & 71.97 & \ding{52} & 3.37 & 12 & 25 & Linear & 1 & \ding{52} & $ 51.27\pm34.43 $ & \ding{53} & -- & \ding{53} & --\\
\rowcolor{rowcolor}
13 & 31 & \ding{52} & 240.24 & \ding{53} & -- & \ding{52} & 39.82 & \ding{52} & 3.07 & 13 & 25 & Linear & 1 & \ding{52} & $ 12.57\pm10.57 $ & \ding{53} & -- & \ding{108} & --\\
14 & 18 & \ding{52} & 79.59 & \ding{53} & -- & \ding{52} & 21.9 & \ding{52} & 3.23 & 14 & 29 & Linear & 1 & \ding{52} & $ 19.42\pm11.52 $ & \ding{53} & -- & \ding{53} & --\\
\rowcolor{rowcolor}
15 & 20 & \ding{52} & 13.66 & \ding{53} & -- & \ding{52} & 274.79 & \ding{52} & 2.43 & 15 & 37 & Linear & 1 & \ding{52} & $ 58.85\pm13.83 $ & \ding{53} & -- & \ding{108} & --\\
16 & 22 & \ding{52} & 30.31 & \ding{53} & -- & \ding{53} & -- & \ding{52} & 6.16 & 16 & 37 & Linear & 1 & \ding{53} & -- & \ding{53} & -- & \ding{108} & --\\
\rowcolor{rowcolor}
17 & 15 & \ding{52} & 24.97 & \ding{53} & -- & \ding{53} & -- & \ding{52} & 2.31 & 17 & 27 & Linear & 1 & \ding{52} & $ 4.80\pm3.62 $ & \ding{52} & 66.4 & \ding{53} & --\\
18 & 22 & \ding{52} & 23.7 & \ding{53} & -- & \ding{53} & -- & \ding{52} & 6.47 & 18 & 24 & Linear & 1 & \ding{52} & $ 19.86\pm12.17 $ & \ding{53} & -- & \ding{53} & --\\
\rowcolor{rowcolor}
19 & 33 & \ding{52} & 21.41 & \ding{53} & -- & \ding{53} & -- & \ding{52} & 2.47 & 19 & 23 & Linear & 1 & \ding{52} & $ 10.34\pm7.59 $ & \ding{53} & -- & \ding{108} & --\\
20 & 22 & \ding{52} & 16.59 & \ding{53} & -- & \ding{52} & 53.57 & \ding{52} & 9.78 & 20 & 25 & Linear & 1 & \ding{53} & -- & \ding{53} & -- & \ding{53} & --\\
\rowcolor{rowcolor}
\multicolumn{2}{c|}{} & \multicolumn{8}{c||}{\cellcolor[HTML]{BFDCDB}113 more cases are omitted due to space limitation} & 21 & 24 & Linear & 1 & \ding{52} & $ 12.09\pm7.97 $ & \ding{53} & -- & \ding{108} & --\\
\cline{1-10}
Total &  & \textbf{114/133} &  & 0/133 &  &73/133 & $ 248\!\pm\!982.2 $ & \textbf{124/133} & $ 2.1\pm2.32$ & 22 & 22 & Linear & 1 & \ding{52} & $ 11.79\pm11.14 $ & \ding{52} & 29.96 & \ding{108} & --\\
\cline{1-10}
\multicolumn{10}{c||}{\normalsize\cellcolor[HTML]{ABF0D8}\textbf{SV-COMP (21 C programs with multiple/nested loops)}} & \cellcolor{rowcolor} 23 & \cellcolor{rowcolor} 28 & \cellcolor{rowcolor} Linear & \cellcolor{rowcolor} 1 & \cellcolor{rowcolor} \ding{52} & \cellcolor{rowcolor} $ 32.71\pm24.14 $ & \cellcolor{rowcolor}\ding{53} & \cellcolor{rowcolor} -- & \cellcolor{rowcolor}\ding{53} & \cellcolor{rowcolor}--\\
\cline{1-10}
\multicolumn{8}{c|}{\textbf{Benchmark Information}} & \multicolumn{2}{c||}{\cellcolor[HTML]{EBCB8B}\textbf{\name}} & 24 & 23 & Linear & 1 & \ding{52} & $ 12.84\pm7.20 $ & \ding{53} & -- & \ding{53} & --\\
\cline{1-10}
\multicolumn{2}{l|}{\textbf{Type}} & \multicolumn{4}{|l}{\textbf{Program}} & \textbf{LoC} & \multicolumn{1}{c|}{\textbf{Loop}} & \textbf{Success} & \textbf{Time(s)} & \cellcolor{rowcolor} 25 & \cellcolor{rowcolor} 62 & \cellcolor{rowcolor} Linear & \cellcolor{rowcolor} 1 & \cellcolor{rowcolor} \ding{52} & \cellcolor{rowcolor} $6.64\pm4.62$ & \cellcolor{rowcolor} \ding{53} & \cellcolor{rowcolor} -- & \cellcolor{rowcolor} \ding{53} & \cellcolor{rowcolor} --\\
\cline{1-10}
\multicolumn{2}{l|}{\multirow{19}{*}{quantifier-free}} & \multicolumn{4}{l}{afnp2014\_true-unreach-call.c} & 17 & \multicolumn{1}{c|}{1} & \ding{52} & $ 40.69\pm34.37 $ & 26 & 32 & Linear & 1 & \ding{52} & $ 44.89\pm49.35 $ & \ding{53} & -- & \ding{53} & --\\
& & \multicolumn{4}{l}{bhmr2007\_true-unreach-call.c} & 31 & \multicolumn{1}{c|}{1} & \ding{52} & $ 4.50\pm2.61 $ & \cellcolor{rowcolor} 27 & \cellcolor{rowcolor} 34 & \cellcolor{rowcolor} Linear & \cellcolor{rowcolor} 1 & \cellcolor{rowcolor} \ding{53} & \cellcolor{rowcolor} -- & \cellcolor{rowcolor} \ding{53} & \cellcolor{rowcolor} -- & \cellcolor{rowcolor} \ding{53} & \cellcolor{rowcolor} --\\
& & \multicolumn{4}{l}{cggmp2005\_true-unreach-call.c} & 18 & \multicolumn{1}{c|}{1} & \ding{52} & $ 128.04\pm53.74 $ & 28 & 26 & Linear & 1 & \ding{52} & $ 11.92\pm9.78 $ & \ding{53} & -- & \ding{53} & --\\
& & \multicolumn{4}{l}{count\_up\_down\_true-unreach-call...} & 22 & \multicolumn{1}{c|}{1} & \ding{52} & $ 18.44\pm15.77 $ & \cellcolor{rowcolor} 29 & \cellcolor{rowcolor} 35 & \cellcolor{rowcolor} Linear & \cellcolor{rowcolor} 1 & \cellcolor{rowcolor} \ding{52} & \cellcolor{rowcolor} $16.21\pm14.83$ & \cellcolor{rowcolor} \ding{53} & \cellcolor{rowcolor} -- & \cellcolor{rowcolor} \ding{53} & \cellcolor{rowcolor} --\\
& & \multicolumn{4}{l}{css2003\_true-unreach-call.c} & 19 & \multicolumn{1}{c|}{1} & \ding{52} & $ 88.07\pm74.35 $ & 30 & 29 & Linear & 1 & \ding{52} & $ 140.51\pm67.06 $ & \ding{53} & -- & \ding{108} & --\\
& & \multicolumn{4}{l}{ddlm2013\_true-unreach-call.c} & 33 & \multicolumn{1}{c|}{1} & \ding{53} & -- & \cellcolor{rowcolor} 31 & \cellcolor{rowcolor} 51 & \cellcolor{rowcolor} Multiple & \cellcolor{rowcolor} 4 & \cellcolor{rowcolor} \ding{53} & \cellcolor{rowcolor} -- & \cellcolor{rowcolor} \ding{53} & \cellcolor{rowcolor} -- & \cellcolor{rowcolor} \ding{53} & \cellcolor{rowcolor} --\\
& & \multicolumn{4}{l}{down\_true-unreach-call.c} & 21 & \multicolumn{1}{c|}{2} & \ding{52} & $ 269.98\pm224.05 $ & 32 & 36 & Multiple & 2 & \ding{52} & $ 71.74\pm60.13 $ & \ding{53} & -- & \ding{53} & --\\
& & \multicolumn{4}{l}{half\_2\_true-unreach-call.c} & 25 & \multicolumn{1}{c|}{2} & \ding{53} & -- & \cellcolor{rowcolor} 33 & \cellcolor{rowcolor} 27 & \cellcolor{rowcolor} Multiple & \cellcolor{rowcolor} 2 & \cellcolor{rowcolor} \ding{52} & \cellcolor{rowcolor} $94.02\pm58.68$ & \cellcolor{rowcolor} \ding{52} & \cellcolor{rowcolor} 26.97 & \cellcolor{rowcolor} \ding{53} & \cellcolor{rowcolor} --\\
& & \multicolumn{4}{l}{hhk2008\_true-unreach-call.c} & 26 & \multicolumn{1}{c|}{1} & \ding{52} & $ 8.49\pm5.87 $ & 34 & 33 & Multiple & 2 & \ding{52} & $ 57.32\pm39.28 $ & \ding{53} & -- & \ding{108} & --\\
& & \multicolumn{4}{l}{jm2006\_variant\_true-unreach-call.c} & 30 & \multicolumn{1}{c|}{1} & \ding{52} & $ 22.52\pm14.55 $ & \cellcolor{rowcolor} 35 & \cellcolor{rowcolor} 27 & \cellcolor{rowcolor} Nested & \cellcolor{rowcolor} 3 & \cellcolor{rowcolor} \ding{52} & \cellcolor{rowcolor} $58.26\pm54.18$ & \cellcolor{rowcolor} \ding{53} & \cellcolor{rowcolor} -- & \cellcolor{rowcolor} \ding{108} & \cellcolor{rowcolor} --\\
& & \multicolumn{4}{l}{jm2006\_true-unreach-call.c} & 25 & \multicolumn{1}{c|}{1} & \ding{52} & $ 123.61\pm100.31 $ & 36 & 30 & Nested & 2 & \ding{52} & $ 84.23\pm57.52 $ & \ding{52} & 17.42 & \ding{53} & --\\
& & \multicolumn{4}{l}{large\_const\_true-unreach-call.c} & 37 & \multicolumn{1}{c|}{2} & \ding{53} & -- & \cellcolor{rowcolor} 37 & \cellcolor{rowcolor} 23 & \cellcolor{rowcolor} Nested & \cellcolor{rowcolor} 2 & \cellcolor{rowcolor} \ding{53} & \cellcolor{rowcolor} -- & \cellcolor{rowcolor} \ding{52} & \cellcolor{rowcolor} 0.26 & \cellcolor{rowcolor} \ding{108} & \cellcolor{rowcolor} --\\
& & \multicolumn{4}{l}{nest-if3\_true-unreach-call.c} & 19 & \multicolumn{1}{c|}{2} & \ding{52} & $ 107.05\pm73.11 $ & 38 & 22 & Nested & 3 & \ding{52} & $ 105.28\pm60.01 $ & \ding{53} & -- & \ding{108} & --\\
& & \multicolumn{4}{l}{nested6\_true-unreach-call.c} & 31 & \multicolumn{1}{c|}{3} & \ding{53} & -- & \cellcolor{rowcolor} 39 & \cellcolor{rowcolor} 35 & \cellcolor{rowcolor} Nested & \cellcolor{rowcolor} 2 & \cellcolor{rowcolor} \ding{52} & \cellcolor{rowcolor} $96.50\pm68.99$ & \cellcolor{rowcolor} \ding{53} & \cellcolor{rowcolor} -- & \cellcolor{rowcolor} \ding{53} & \cellcolor{rowcolor} --\\
& & \multicolumn{4}{l}{nested9\_true-unreach-call.c} & 23 & \multicolumn{1}{c|}{3} & \ding{52} & $274.87\pm156.77$ & 40 & 26 & Nested & 3 & \ding{52} & $ 139.88\pm92.53 $ & \ding{52} & 147.89 & \ding{53} & --\\
& & \multicolumn{4}{l}{seq\_true-unreach-call.c} & 33 & \multicolumn{1}{c|}{3} & \ding{53} & -- & \cellcolor{rowcolor} 41 & \cellcolor{rowcolor} 30 & \cellcolor{rowcolor} Nested & \cellcolor{rowcolor} 2 & \cellcolor{rowcolor} \ding{52} & \cellcolor{rowcolor} $177.43\pm106.79$ & \cellcolor{rowcolor} \ding{53} & \cellcolor{rowcolor} -- & \cellcolor{rowcolor} \ding{53} & \cellcolor{rowcolor} -- \\
& & \multicolumn{4}{l}{sum01\_true-unreach-call...} & 20 & \multicolumn{1}{c|}{1} & \ding{52} & $8.12\pm6.41$ & 42 & 29 & Nested & 3 & \ding{52} & $228.71\pm218.43$ & \ding{53} & -- & \ding{53} & --\\
& & \multicolumn{4}{l}{terminator\_03\_true-unreach-call...} & 20 & \multicolumn{1}{c|}{1} & \ding{52} & $ 26.09\pm22.98 $ & \cellcolor{rowcolor} 43 & \cellcolor{rowcolor} 39 & \cellcolor{rowcolor} Nested & \cellcolor{rowcolor} 3 & \cellcolor{rowcolor} \ding{53} & \cellcolor{rowcolor} -- & \cellcolor{rowcolor} \ding{53} & \cellcolor{rowcolor} -- & \cellcolor{rowcolor} \ding{53} & \cellcolor{rowcolor} --\\
& & \multicolumn{4}{l}{up\_true-unreach-call.c} & 30 & \multicolumn{1}{c|}{2} & \ding{52} & $ 374.38\pm291.44 $ & 44 & 61 & Nested & 4 & \ding{53} & -- & \ding{53} & -- & \ding{53} & --\\
\cline{1-10}
\multicolumn{2}{l|}{\multirow{2}{*}{quantifier}} & \multicolumn{4}{l}{array\_true-unreach-call1.c} & 13 & \multicolumn{1}{c|}{1} & \ding{52} & $ 5.88\pm3.78 $ & \cellcolor{rowcolor} 45 & \cellcolor{rowcolor} 46 & \cellcolor{rowcolor} Nested & \cellcolor{rowcolor} 3 & \cellcolor{rowcolor} \ding{52} & \cellcolor{rowcolor} $202.78\pm185.39$ & \cellcolor{rowcolor} \ding{53} & \cellcolor{rowcolor} -- & \cellcolor{rowcolor} \ding{53} & \cellcolor{rowcolor} --\\
& & \multicolumn{4}{l}{array\_true-unreach-call2.c} & 18 & \multicolumn{1}{c|}{1} & \ding{52} & $ 32.53\pm27.11 $ & 46 & 24 & Nested & 3 & \ding{52} & $ 86.61\pm50.22 $ & \ding{53} & -- & \ding{108} & --\\
\midrule
\multicolumn{2}{l|}{\textbf{Total}} & \multicolumn{6}{c|}{} & \textbf{16/21} & \multicolumn{1}{r||}{\textbf{$217.9$}\,$\pm$\,$451.7$} & & & & & \textbf{38/46} & $ 68.92\pm171.13 $  & 9/46 & $ 83.0\pm104.8 $ & 0/46 & \\
\bottomrule
\end{tabular}
}
\vspace{-10pt}
\end{table*}

\subsection{RQ2. Effectiveness on Loop Invariants \iffalse
Specification\fi}\label{sec:rq2}
Table~\ref{tab:rq2-merge} shows the effectiveness of \name in generating specifications of loop invariants compared with three baselines. In particular, the SyGuS benchmark consists of 133 C programs. Each program contains only one loop structure. We compare \name with three baselines: Pilat~\cite{de2016polynomial}, Code2Inv~\cite{code2inv18} and CLN2INV~\cite{ryan2020cln2inv} on this benchmark. The result is shown in Table~\ref{tab:rq2-merge} under \textit{SyGuS} entry. Pailt fails to generate valid specifications for all cases in this benchmark, as all the specifications it generates are either unsatisfiable or irrelevant. On the other hand, Code2Inv and CLN2INV perform better, solving 73 and 124 programs, respectively. \name can handle a comparable number of cases, namely, 114 programs in this benchmark. Although CLN2INV can solve 10 more cases in this benchmark, it cannot handle any cases in the OOPSLA-13 benchmark. Although CLN2INV can successfully parse 19 out of 46 cases (denoted as {\tiny \ding{108}}), \iffalse after parsing,\fi CLN2INV fails to construct \iffalse valid\fi satisfiable invariants that are adequate to \iffalse prove\fi verify the programs, resulting in a score of 0/46. This could be due to the overfitting of machine learning methods to specific datasets. Code2Inv, on the other hand, can handle 9 out of 46 cases. In comparison, \name can solve 38/46 (82.60\%), which significantly outperforms existing approaches.

Furthermore, we consider a more difficult benchmark, SV-COMP.
Due to the unsatisfactory results of the existing approaches, we have opted to exclusively present the results obtained using \name. 
As shown in Table~\ref{tab:rq2-merge} under \textit{SV-COMP} entry, \name can solve 16 out of 21 programs{\iffalse in the benchmark\fi} with an average time of 3 minutes. Note that there are three programs with 3-fold nested loop structures in this benchmark\iffalse, \ie, \texttt{nested6\_true-unreach-}\texttt{call.c}, \texttt{nested9\_true-unreach-call.c} and \texttt{seq\_true-unreach-} \texttt{call.c}\fi. \name can solve one of them\iffalse (\ie, \texttt{nested9\_true-unreach-call.c})\fi, while for the other \iffalse rest\fi two programs, \iffalse though\fi \name can generate several \iffalse correct\fi satisfiable specifications, but there are still one or two specifications that cannot be generated after five iterations.

\subsection{RQ3. Efficiency of \name}\label{sec:rq3}

\begin{figure*} [t]
    \centering
    \vspace{-0ex}
    \includegraphics[width=1.0\linewidth]{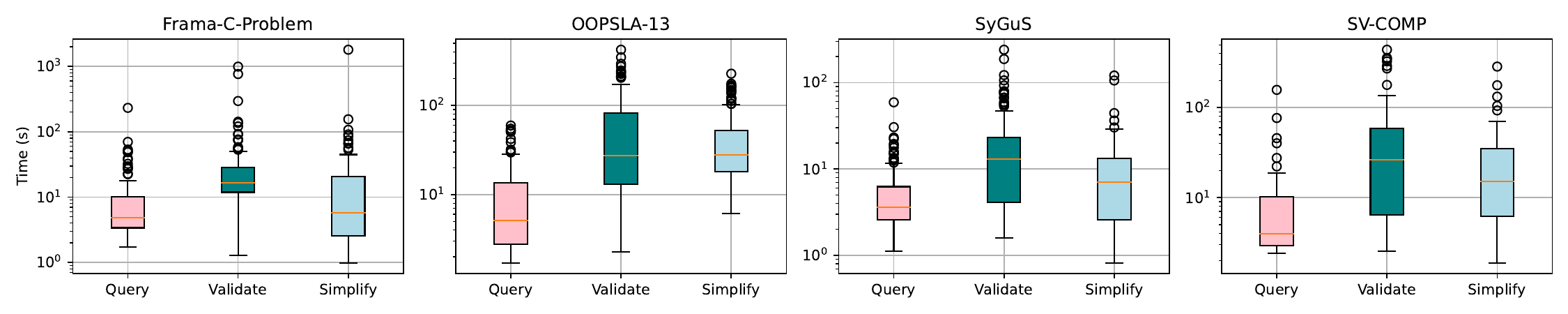}
    \setlength{\abovecaptionskip}{-10pt}
    \setlength{\belowcaptionskip}{-10pt}
    \caption{\textbf{Overhead of \name on Four Benchmarks}. {\iffalse{\footnotesize\textit{The time in seconds used in querying LLMs (Query), validating and verifying using verification tool (Validate) and simplifying the generated specifications (Simplify) on the four benchmarks \iffalse\framac-problem~\cite{frama-c-problems}, OOPSLA-13~\cite{oopsla13}, SyGuS~\cite{SyGuS} and SV-COMP~\cite{SVCOMP}\fi are visualized.}}\fi}}
    \label{fig:overhead}
\end{figure*}

In the first RQs, we can observe that \name can generate satisfiable and adequate specifications for the proof ranging from 2.35 to 739.62 seconds (\ie, 0.04 to 12.33 minutes). 
In this RQ, we illustrate the composition of the overhead in \name across sub-tasks, \ie, the time required for querying the LLM for specifications, validating and verifying the specifications against the theorem prover, and simplifying the specifications \iffalse simplification \fi(optional). The results for these four benchmarks are presented in Figure~\ref{fig:overhead}.

We can see that \iffalse across\fi for the four benchmarks, validating and verifying the specifications (\textit{Validate}) takes the most time, ranging from 1.3 to 994.9 seconds. Querying the LLMs (\textit{Query}) takes the least time, \iffalse with less than 10 seconds on average\fi averaging less than 10 seconds. 
It is noteworthy that, unlike existing works that tend to generate a lot of candidate specifications and check their validity for one hour~\cite{ryan2020cln2inv} to 12 hours~\cite{code2inv18}, 
\name \iffalse spent\fi takes far less time in validating (\eg, 1.2 seconds to 3.88 minutes). This is because \name generates fewer but higher-quality specifications. The efficiency of \name makes it both practical and cost-effective for various applications.

In addition, the time required for simplifying the specifications (\textit{Simplify}) may vary depending on the number of generated specifications. A larger number of specifications leads to a longer simplification process. Nonetheless, given the fact that the simplification step is optional in \name, and considering the benefit of faster solving brought about by the concise and elegant specifications, the cost of simplification is justified.

\subsection{RQ4. Ablation Study}\label{sec:rq4}
Finally, we evaluate the contribution made by each part of \name's design. The results are shown in Table~\ref{tab:rq4}. We conduct the evaluation on \framac-problems benchmark~\cite{frama-c-problems} under seven settings: (1) - (4) settings under \textit{Base ChatGPT} entry directly feed the C program together with the desired properties to be verified into ChatGPT{\iffalse (\textit{gpt-3.5-turbo-0613})\fi}, with {zero-}/{one-}/{two-}/{three-shot}. These settings are designed to compare with the decomposed manner adopted by \name. Setting (5) under entry \textit{Decomposed} adopts the code decomposition (\ie, Step 1 of \name)  with three-shot,  because it shows the best result according to the results of the previous settings. Settings (6) and (7) are respectively configured with only one pass (\ie, without enhancement) and five iterations (\ie, with enhancement), showing the improvement brought by the iterative enhancement (Step (K) in Figure~\ref{fig:overview}). The last row shows the total number of programs that can be successfully solved under the corresponding settings. 

Table~\ref{tab:rq4} shows an ascending trend in the number of solved programs, from 5 to 31 over 51. On the one hand, it is hardly possible to directly ask ChatGPT to generate specifications for the entire program. \iffalse can hardly succeed.\fi The input-output examples\iffalse,  though slightly improving the result (from 5 to 9),\fi ~bring only a limited improvement (from 5 to 9) in the performance. On the other hand, \iffalse decomposing the programs and asking ChatGPT to generate specifications hierarchically\fi code decomposition and hierarchical specification
generation bring \iffalse an obvious\fi a significant improvement (from 9 to 26). This shows the contribution made by the first two steps of \name. Furthermore, the contribution of iterative enhancement can be observed in the last two columns, from 27 to 31. Overall, the ablation study shows that every step in \name has a positive \iffalse effect\fi impact on the final result and that the idea of code decomposition and hierarchical specification
generation \iffalse contribute the most\fi brings the biggest improvement.

\begin{table}[t!]
\centering
\setlength{\abovecaptionskip}{-0pt}
\setlength\tabcolsep{2pt}
\setlength{\belowcaptionskip}{-10pt}
\caption{Experiment Result of Ablation Study}
\label{tab:rq4}
\scriptsize
\renewcommand\arraystretch{1.0}
    \resizebox{1.0\linewidth}{!}{
\begin{tabular}{l|cccc|c|cc}
\toprule
 \multirow{2}{*}{\textbf{Type}} & \multicolumn{4}{c|}{\textbf{Base ChatGPT}} & \multicolumn{1}{c|}{\textbf{Decomposition}} & \multicolumn{2}{c}{\begin{tabular}[c]{@{}c@{}}\textbf{Iterative}\\ \textbf{Enhancement}\end{tabular}} \\ 
  \cmidrule{2-8} 
 & \multicolumn{1}{c}{(1) 0-shot} & \multicolumn{1}{c}{(2) 1-shot} & \multicolumn{1}{c}{(3) 2-shot} & \multicolumn{1}{c|}{(4) 3-shot} & \multicolumn{1}{c|}{(5) 3-shot} & \multicolumn{1}{c}{(6) Pass@1} & \multicolumn{1}{c}{(7) Iter@5} \\ 

 \midrule
Loops & 1 & 1 & 1 & 1 & 1 & 1 & 1 \\
Immutable\_arrays & 0 & 0 & 0 & 0 & 3 & 4 & 5 \\
Mutable\_arrays & 0 & 0 & 0 & 0 & 0 & 0 & 2 \\
Arrays\_and\_loops & 1 & 0 & 1 & 1 & 4 & 4 & 4 \\
More\_arrays & 0 & 0 & 0 & 0 & 1 & 1 & 1 \\
General\_wp\_problems & 2 & 4 & 3 & 5 & 8 & 8 & 8 \\
Pointers & 0 & 2 & 0 & 1 & 6 & 6 & 7 \\
Miscellaneous & 1 & 1 & 1 & 1 & 3 & 3 & 3 \\
\midrule
\textbf{Total} & 5 & 8 & 6 & 9 & 26 & 27 & \textbf{31}\\
\bottomrule
\end{tabular}
}
\vspace{-5pt}
\end{table}

\subsection{Case Studies} \label{sec:case-study}

We discuss two representative cases to show how iterative enhancement contributes (Fig.~\ref{listing:case3}), and a situation where \name fails to handle (Fig.~\ref{listing:case2}).

\smallskip
\noindent{\textbf{Case 1. A success case made by validation and iterative Enhancement}.
We show how specification validation and iterative enhancement help \name to generate \iffalse strong\fi satisfiable and adequate specifications. The program presented below computes the sum of three values stored in pointers.
In the first iteration, only two specifications (lines~\ref{line:spec1} and \ref{line:spec2}) are generated, which respectively require three pointers should be valid (line~\ref{line:spec1}), and the result of \texttt{add} is the sum \iffalse summation\fi (line~\ref{line:spec2}). However, these two specifications alone are \iffalse insufficient\fi inadequate to \iffalse prove\fi verify the property due to the lack of a specification describing whether the values of the pointers have been modified within the \texttt{add} function.
\name then inserts placeholders immediately after the two generated specifications and continues \iffalse proceeds\fi to the second iteration of enhancement. The subsequently generated specification (line~\ref{line:spec3}) states the \texttt{add} function has no assignment behavior, \iffalse which is subsequently generated, \fi making the verification succeed.

\smallskip
\noindent{\textbf{Case 2. A failing case due to missing context}}.
We present an example where \name fails to generate adequate specifications due to \textit{the lack of necessary context}. The code for the \texttt{pow} in \texttt{<math.h>} is not directly accessible. Currently, \name does not automatically trace all the dependencies and include their code in the prompt. LLMs can hardly figure out what \texttt{pow} is expected to do. As a result, the specification for this function cannot be generated \iffalse after\fi despite all attempts made by \name. This case shows a possible \iffalse potential\fi improvement by adding \iffalse including\fi more dependencies in the prompt.
\smallskip

\definecolor{stubbg}{HTML}{FCF3D5}
\begin{figure}[t]
\begin{lstlisting}[
		language=c,
        basicstyle=\linespread{1.25}\tiny,
		morekeywords={var},
        aboveskip=15pt,
        belowskip=0pt,
		%label={listing:case3},
		escapechar=|,
        captionpos=none,
		linebackgroundcolor = {\ifnum \value{lstnumber} > 3 \ifnum \value{lstnumber} < 5 \color{stubbg} \fi \fi},
		numbers=left
	]
/*@
requires|\;|\|\!|valid(|\!|a|\!|)|\;|&&|\;|\|\!|valid(|\!|b|\!|)|\;|&&|\;|\|\!|valid(|\!|r|\!|)|\!|; |\label{line:spec1}|
ensures|\,|\result == *a + *b + *r;|\label{line:spec2}|
assigns|\,|\nothing;|\label{line:spec3}|
*/
int add(int *a, int *b, int *r)
{    
  return *a + *b + *r; 

}
\end{lstlisting}
    \setlength{\abovecaptionskip}{5pt}
    \setlength{\belowcaptionskip}{-10pt}
    \caption{Case 1}
    \label{listing:case3}
\end{figure}
\definecolor{stubbg}{HTML}{FCF3D5}
\begin{figure}[t]
\begin{lstlisting}[
		language=c,
        basicstyle=\linespread{1.25}\tiny,
		morekeywords={var},
        aboveskip=15pt,
        belowskip=0pt,
		%label={listing:case2},
		escapechar=|,
        captionpos=none,
		linebackgroundcolor = {\ifnum \value{lstnumber} > 5 \ifnum \value{lstnumber} < 7 \color{stubbg} \fi \fi},
		numbers=left
	]
#include <math.h>
int fun(int n) {
  double y = 0, i = 0;
  /*@ loop|\,|invariant y == (\pow(2,|\,|i))|\,|-|\,|1;
      loop|\,|invariant i <= n; */
  while(i <= n){
    y = y + pow(2.0, i);
    i = i + 1; }
  return y;
}
\end{lstlisting}
    \setlength{\abovecaptionskip}{5pt}
    \setlength{\belowcaptionskip}{-10pt}
    \caption{Case 2}
    \label{listing:case2}
\end{figure}

%%%%%%%%%%%%%%%%%%%%%%%%%%%%%%%%%%%%%%%%%%%%%%%%%%
%%%%%%%%%%%%%%%% Threats to Validity %%%%%%%%%%%%%
%%%%%%%%%%%%%%%%%%%%%%%%%%%%%%%%%%%%%%%%%%%%%%%%%%
\section{Threats to Validity}\label{sec:threat}
There are three major validity threats. The first concerns \textbf{the data leakage problem}. We addressed this threat in two folds. First, we directly apply LLMs to generate the specifications (Section~\ref{sec:rq4}). The unsatisfactory result (success rate: 5/51) shows that the chance of overfitting to the benchmark is low. Second, we followed a recent practice~\cite{WuJPLD0BS23} for the data leakage threat. We randomly sampled 100 programs from three benchmarks in RQ2 (\ie, SyGuS, OOPSLA-13, and SV-COMP) in a ratio of 50:25:25 and mutated these programs by variable renaming (\eg, renaming \texttt{x} to \texttt{m}) and statement/branch switching (\eg, negotiating the if-condition, and switching the statements in \texttt{if} and \texttt{else} branches) without changing the semantics of the program manually. Then we applied \name over the 100 mutated programs. The experiment shows that \textit{98\% results hold after the programs are mutated}. It further confirmed that the validity threat of data leakage is low. 
The second concerns \textbf{the generalizability to different LLMs}. To address this concern, we implemented \name to a popular and open-source LLM called \textit{Llama-2-70b} and ran it on the same benchmark used in RQ1. Similar results were observed, with \name (Llama2) achieving a score of 25/51 compared to the score of 31/51 achieved by \name (ChatGPT).
The third concerns \textbf{the scalability of \name}. We have evaluated a real-world X509-parser project and achieved unexpectedly good performance. However, completing the whole verification task on the entire project remains challenging. The evidence suggests that \name has the potential to assist participants in writing specifications for real-world programs.

\section{Related Work}\label{sec:related}

\noindent\textbf{Specification Synthesis.}
While there exist various approaches and techniques for generating program specifications from natural language~\cite{nl2spec,zhai2020c2s,giannakopoulou2020generation}, this paper primarily focuses on specification generation based on the programming language.
There has been work using data mining to infer specifications~\cite{beckman2011probabilistic,lo2007efficient,le2018deep,kang2021adversarial}.
Several of these techniques use dynamic traces to infer possible invariants and preconditions from test cases, and static analysis to check the validity and completeness of the inferred specifications~\cite{ammons2002mining,yang2006perracotta,nimmer2002automatic}.
While others apply domain knowledge and statically infer specifications from the source code~\cite{ramanathan2007static,beckman2011probabilistic,shoham2007static}. 
Several works have been conducted to address the challenging sub-problem of loop invariant inference, including CLN2INV~\cite{ryan2020cln2inv}, Code2Inv~\cite{code2inv18}, G-CLN~\cite{yao2020learning} and Fib~\cite{fib17}. Additionally, there are also studies dedicated to termination specification inference~\cite{le2015termination}.
A recent study, SpecFuzzer~\cite{molina2022fuzzing}, combines grammar-based fuzzing, dynamic invariant detection, and mutation analysis to generate class specifications for Java methods in an automated manner.
Our approach differs from these techniques as it statically generates comprehensive contracts for each loop and function, yielding reliable outcomes necessary for verification.

\smallskip
\noindent\textbf{Assisting Program Analysis and Verification with LLMs.}
In recent years, there has been a growing interest in applying LLMs to assist program analysis tasks~\cite{llmse}, such as fuzz testing~\cite{yinlin-issta23,codamosa-icse23}, static analysis~\cite{wen2024automatically,Weisong-CoRR23,Haonan-sp23}, program verification~\cite{first2023baldur,wu2023lemur,yang2023leandojo}, bug reproduction~\cite{Sungmin-icse23} and bug repair~\cite{Hammond-sp23,Chunqiu-icse23,Zhiyu-icse23}. 
For example, Baldur~\cite{first2023baldur} is a proof-synthesis tool that uses transformer-based pre-trained large language models fine-tuned on proofs to generate and repair whole proofs. 
In contrast, {\name} focuses on generating various types of program specifications and leveraging the auto-active verification tool to complete the verification task, while Baldur focuses on automatically generating proofs for the theorems.
Li \textit{et al.}~\cite{Haonan-sp23} investigated the potential of LLMs in enhancing static analysis by posing relevant queries. 
They specifically focused on UBITest~\cite{ubitest-fse20}, a bug-finding tool for detecting use-before-initialization bugs. 
The study revealed that those false positives can be significantly reduced by asking precisely crafted questions related to function-level behaviors or summaries.
Ma \textit{et al.}~\cite{ma-CoRR23} and Sun \textit{et al.}~\cite{Weisong-CoRR23} explore the capabilities of LLMs when performing various program analysis tasks such as control flow graph construction, call graph analysis, and code summarization.
Pei \textit{et al.}~\cite{pei2023can} use LLMs to reason about program invariants with decent performance.
These diverse applications underline the vast potential of LLMs in program analysis.
{\name} complements these efforts by showcasing the effectiveness of LLMs in generating practical and elegant program specifications, thereby enabling complete automation of deductive verification.

\iffalse
\noindent\textbf{Automation of Deductive Verification.}
Automating program verification is a critical challenge in modern software development. To address this challenge, automation of two specific problems is essential: specification generation/elimination and theorem proving for verification conditions.
In the last decade, significant efforts have been dedicated to interactive and automated theorem proving~\cite{loveland2016automated}, leading to the development of various auto-active verification tools~\cite{blanchard2019towards,tschannen2015autoproof}.
Though there are existing studies that utilize a symbolic method called loop invariant elimination~\cite{kondratyev2019towards,kondratyev2022automation}, these approaches are often limited in their application to linear arrays.
Our approach focuses on specification generation, which boasts greater versatility and applicability across diverse contexts.
\fi

\section{Conclusion}\label{sec:conclu}
In this paper, we presented {\name}, a novel approach for generating program specifications from source code. Our approach leverages the power of Large Language Models (LLMs) to infer the candidate program specifications in a bottom-up manner, and then validates them using provers/verification tools and iteratively enhances them.
The evaluation results demonstrate that our approach to specification generation achieves full automation and cost-effectiveness, which is a major bottleneck for formal verification.

\vspace{-0.5ex}
\section{Data Availability}
\vspace{-0.5ex}
We have released the implementation and all associated publicly available data:
{\color{blue}\url{https://sites.google.com/view/autospecification}}.

\section*{Acknowledgements}
The authors would like to thank the anonymous reviewers for their constructive comments.
This work was supported in part by the National Natural Science Foundation of China (Nos. 62372304, 62302375, 62192734), the China Postdoctoral Science Foundation funded project (No. 2023M723736), and the Fundamental Research Funds for the Central Universities.

%\clearpage

%\newpage

%\balance
%\bibliographystyle{ACM-Reference-Format}
\bibliographystyle{unsrt} %splncs
\bibliography{Tex/reference}

%\clearpage

%\appendix
%\section{Appendix}\label{sec:appendix}
%\input{appendix}

\end{document}